\pdfoutput=1 \pdfsuppresswarningpagegroup=1 
\documentclass[
 aps,% 
 prb,%
 reprint, twocolumn,%
 groupedaddress,% 
 superscriptaddress,%
 amsfonts,%
 letterpaper,%
 footinbib,%
 floatfix,%
 longbibliography
]{revtex4-2}

\RequirePackage{amsmath,amssymb,bm}
\RequirePackage{graphicx}
\RequirePackage[usenames,dvipsnames]{xcolor}
\usepackage{float}

\usepackage[T1]{fontenc}

\usepackage{enumitem}

\usepackage{verbatim}

\usepackage{hyperref}
\hypersetup{
    colorlinks=true,
    citecolor= blue,
    linkcolor= magenta,
    filecolor=magenta,      
    urlcolor= blue,
    pdfauthor = {Chuan Chen},
    pdfcreator = {\LaTeX\ and \flqq hyperref\frqq},
}

\usepackage[capitalise]{cleveref}

\graphicspath{ {./Figs/} }

\newcommand{\Fref}[1]{Fig.~\ref{#1}}
\newcommand{\Frefs}[1]{Figs.~\ref{#1}}
\newcommand{\Eqref}[1]{Eq.~\eqref{#1}}

\newcommand{\Secref}[1]{Sec.~\ref{#1}}

\newcommand{\Appref}[1]{Appendix~\ref{#1}}

\newcommand{\bA}{\mathbf{A}}
\newcommand{\bB}{\mathbf{B}}
\newcommand{\pp}{\mathsf{p}}
\newcommand{\vv}{\mathsf{v}}

\newcommand{\SW}{\text{SW}}

\newcommand{\bR}{\mathbf{R}}
\newcommand{\phiut}{\phi_\text{UT}}
\newcommand{\philt}{\phi_\text{LT}}
\newcommand{\phiuc}{\phi_\text{UC}}

\makeatletter
\newcommand*{\rom}[1]{\expandafter\@slowromancap\romannumeral #1@}
\makeatother

\renewcommand{\Im}{\operatorname{Im}}

\newcommand{\teq}{{\,=\,}}

\newcommand{\tlt}{{\,<\,}}

\newcommand{\tle}{{\,\le\,}}

\DeclareMathOperator{\sgn}{sgn}

\begin{document}

\title{The nature of visons in the perturbed ferromagnetic and antiferromagnetic
Kitaev honeycomb models}

\author{Chuan~Chen}
\email{chuanchen@tsinghua.edu.cn}
\affiliation{
Institute for Advanced Study, Tsinghua University,
100084 Beijing, China
}
% \affiliation{
% Max-Planck Institute for the Physics of Complex Systems,
% 01187 Dresden, Germany
% }

\author{Inti Sodemann Villadiego}
\email{sodemann@itp.uni-leipzig.de}
\affiliation{Institut f\"ur Theoretische Physik,
Universit\"at Leipzig,
04103 Leipzig, Germany
}
\affiliation{
Max-Planck Institute for the Physics of Complex Systems,
01187 Dresden, Germany
}

\begin{abstract}
The Kitaev honeycomb model hosts a fascinating fractionalized state of matter
featuring emergent Majorana fermions and a vison particle that carries the flux of
an emergent gauge field.
In the exactly solvable model these visons are static but certain perturbations can
induce their motion. We show that the nature of the vison motion induced by
a Zeeman field is sharply distinct in the ferromagnetic vs the
antiferromagnetic Kitaev models.
Namely, in the ferromagnetic model the vison has a trivial non-projective
translational symmetry, whereas in the antiferromagnetic Kitaev model it has a
projective translational symmetry with $\pi$-flux per unit cell. The vison band of the ferromagnetic case has zero Berry curvature,
and no associated intrinsic contribution to the thermal Hall effect.
In contrast, in the antiferromagnetic case there are two
gapped vison bands with opposite Chern numbers and an associated intrinsic vison
contribution to the thermal Hall effect.
We discuss these findings in the light of the physics of the spin liquid
candidate $\alpha$-RuCl$_3$.    
\end{abstract}

\date{\today}

\maketitle

\section{Introduction}
The Kitaev honeycomb model~\cite{Kitaev2006} has become a paradigmatic playground to
investigate spin liquids with emergent fermions and $Z_2$ gauge fields in two-dimensions.
Unlike other exactly solvable models, the model only contains spin bilinears and
it is believed to be a reasonable description of certain quantum magnets,
such as $\alpha$-RuCl$_3$~\cite{jackeli2009mott,winter2017models,takagi2019concept}.
Although $\alpha$-RuCl$_3$ forms a zig-zag antiferromagnet in the absence of applied magnetic fields~\cite{johnson2015monoclinic,cao2016low,banerjee2018excitations},
there are several experimental indications that it might harbor a quantum spin
liquid once an in-plane magnetic field is applied within a range of
$\sim 6T - 11T$~\cite{kasahara2018unusual,banerjee2016proximate,sandilands2015scattering,nasu2016fermionic,yoshitake2016fractional,suetsugu2022evidence,kasahara2018majorana,yokoi2021half,bruin2022robustness,tanaka2022thermodynamic,czajka2021oscillations}.

Nevertheless, the connection between the experimentally observed potential
spin-liquid in $\alpha$-RuCl$_3$ and the spin liquids realized in the weakly perturbed
ideal Kitaev model is currently far from clear. This stems in part from the lingering
uncertainty on the minimal Heisenberg-type model describing the material.
The largest coupling term in $\alpha$-RuCl$_3$, denoted by $K$, is in fact believed
to be the term that appears in the ideal Kitaev model. While a majority studies have
advocated that this coupling is ferromagnetic (FM, $K<0$), others have advocated for
an antiferromagnetic (AFM, $K>0$) exchange coupling
(see Refs.~\cite{laurell2020dynamical,maksimov2020rethinking} for tables summarizing
the estimates of several studies).
Some of the prominent observational evidence favoring $K$ to be ferromagnetic come
from elastic X-ray scattering experiments~\cite{Sears2020} that determined the
direction of the ordered moment in the zig-zag AFM state, which is dependent on the
sign of $K$~\cite{Chaloupka2016}. Inelastic X-ray scattering experiments~\cite{Suzuki2021}
have also advocated for a ferromagnetic coupling. However, these experiments relied
on modeling the zig-zag AFM state, which is highly susceptible to perturbations and
itself subjected to very strong quantum fluctuations. Therefore we believe that
there is still some room to be reasonably skeptic about the certainty of the sign
of this coupling. Determining this sign is crucial for many reasons.
For example the spin liquid realized in the AFM coupled case is more robust to
certain perturbations, as compared to the spin liquid realized for the FM coupled case~\cite{hickey2019emergence,zhu2018robust,gordon2019theory,gohlke2018dynamical}.
Additionally, the nature of the phases driven by the applied Zeeman field can be
very different in both cases, displaying a delicate dependence on the further-neighbor
exchange couplings allowed by symmetry, such as the $\Gamma,\Gamma',J,J_3$
terms~\cite{Winter_2017,rau2014trigonal,hickey2019emergence,zhu2018robust,gordon2019theory,gohlke2018dynamical,laurell2020dynamical,maksimov2020rethinking,kim2015kitaev,sorensen2021heart,gordon2019theory}.

In the present study we will further emphasize the importance of the sign of
$K$ by demonstrating that the FM Kitaev model is in a sharply distinct
topological phase compared to the AFM Kitaev model in the presence of a Zeeman field.
More specifically, we will show that even though the FM and AFM Kitaev models
realize ground states within the same celebrated Ising topological order,
they realize distinct symmetry enriched topological orders with regard to the
translational symmetry of the lattice, and therefore belong to two distinct
universality classes. This distinction manifests vividly on the properties of
its vison quasiparticle~\cite{Read1989,Kivelson1989,Read1991,Senthil2000},
which is the emergent non-abelian anyon analogous to a $\pi$-flux in a $p+ip$
superconductor that carries a Majorana fermion zero mode in its core~\cite{Read2000}.

%\cc{In this work, we will be focusing on the model containing both Kitaev
%couplings and a finite Haldane mass
%($\kappa$ term in \Eqref{eq:H_K}), which can always be generated perturbatively
%in the presence of an external Zeeman field.
%Moreover, as the spinon excitation is gapped with a finite $\kappa$,
%it makes the vison hopping a well-defined quantity as will be discussed
%below.
%Therefore, unless specified explicitly, below we shall denote the model
%with a non-vanishing mass ($\kappa \neq 0$ in \Eqref{eq:H_K}) as Kitaev model.}

\begin{figure}
\centering
\includegraphics[width=0.48 \textwidth]{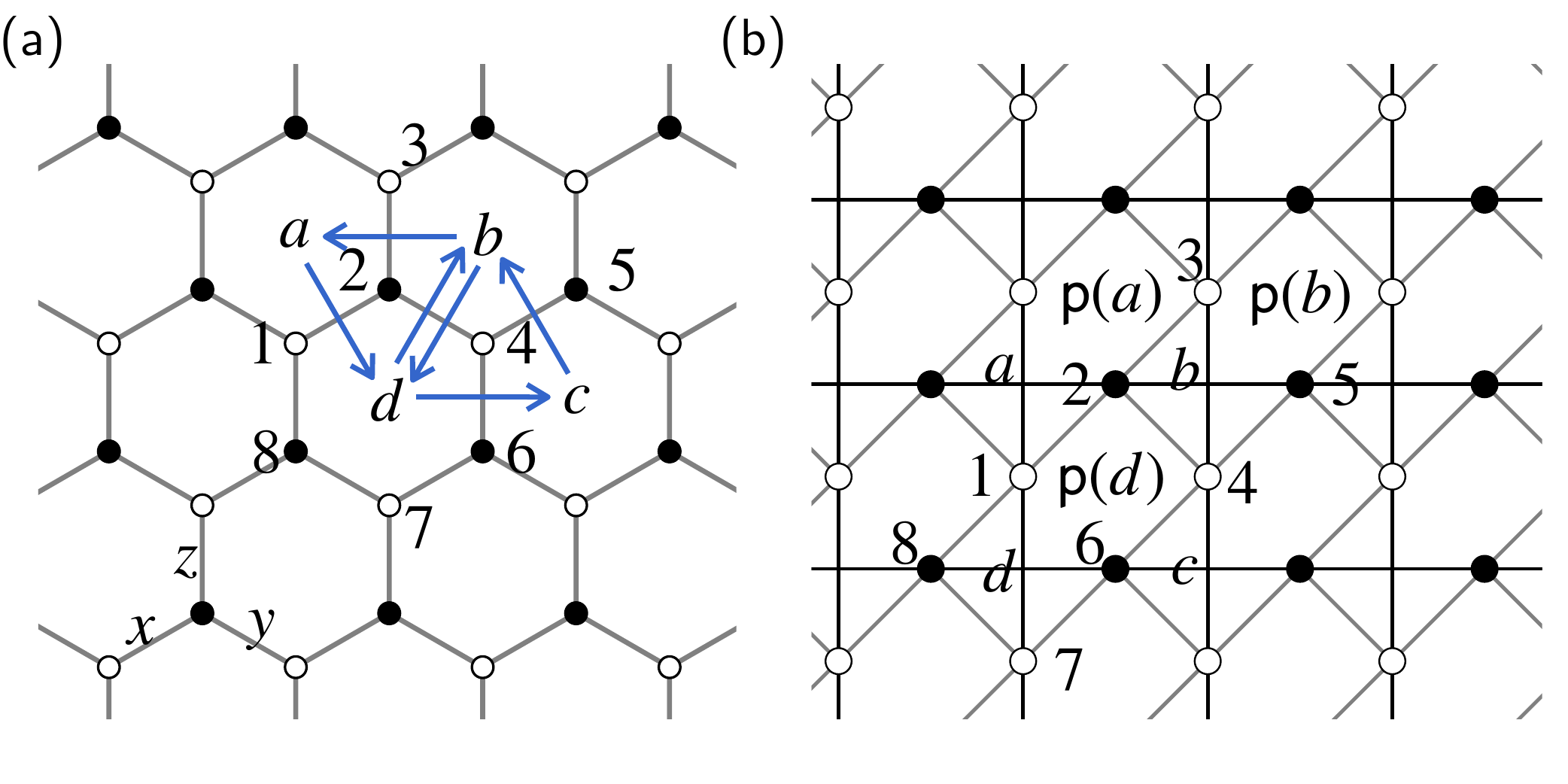}
\caption{
(a) Honeycomb lattice of the Kitaev model. The $x$-, $y$- and $z$-links are
labeled respectively.
The visons are located at the centre of hexagons, e.g.,
$a$, $b$, $c$ and $d$ highlighted in the schematic.
$A$/$B$-sublattice sites are represented by open/filled disks.
(b) The tilted honeycomb lattice with vertices placed on the links of a
square lattice. Visons are now located at the vertices of the
square lattice, while the $\varepsilon$ fermions sit at plaquettes.
$A$-sublattice sites of the original honeycomb lattice now all align on
vertical links of the square lattice.
}\label{fig:schematic}
\end{figure}

We will show that for the FM Kitaev model, lattice translations act on the vison in
an ordinary non-projective way. This implies that the vison Bravais lattice contains
a single state per unit cell associated with each hexagon of the honeycomb lattice
(see \Fref{fig:schematic}(a)), and as a consequence its Berry curvature vanishes
everywhere in its Brillouin zone. 
One important aspect that we will emphasize in our study is that in order to
correctly capture the motion of the vison induced by the Zeeman field, 
it is crucial to include the Haldane mass term
that is generated perturbatively by the Zeeman field on the itinerant Majorana fermions.
Such a term is strictly necessary in order to make the state a fully
gapped topologically ordered phase of matter and to make the vison an exponentially
localized particle. 
In fact, we will directly show numerically that in finite
size systems the phase that the vison acquires is highly sensitive on system
size and boundary conditions when the Haldane mass of the Majorana is set to be
exactly zero. On the other hand, we will also show numerically that as the
thermodynamic limit is approached any small but non-zero Haldane mass term is
sufficient in order to regularize and obtain a unique and well defined phase
for the visons as they moves around a unit cell of the Bravais lattice. 

On the other hand, we will show that for the AFM Kitaev model the presence of the Haldane mass term leads to a finite vison hopping. We will show that interestingly in the AFM case the vison has
indeed projective translations with $\pi$-flux per unit
cell allowing it to have a finite  Berry curvature and two Chern bands with Chern
numbers $C \teq \pm 1$.
% ~\footnote{This case was not studied in Ref.~\cite{Joy2021} because that study performed calculations in the case of strictly zero Haldane mass term, and in this case the vison hopping in the AFM case vanishes in the thermodynamic limit to leading order in the perturbing Zeeman field.}.
We will also show, in agreement with Ref.~\cite{Joy2021},
that perturbatively in the applied Zeeman field, the vison in the FM Kitaev model
has a larger band-width than in the AFM Kitaev model,
so the vison band minimum reaches zero faster in the
former (FM) case as the Zeeman field increases (see \Fref{fig:vison-bands} for a plot of
vison bands in both FM and AFM Kitaev models).
This is crucially related to the stability of the latter against Zeeman perturbations that
would stabilize competing ordered phases, because some of these instabilities could
be viewed as gauge confinement transitions driven by vison condensation that would
appear as its band-width increases and the vison gap closes at certain momenta.

Our work is also interesting from the point of view of the mathematical methods that we exploit to compute these properties. In fact our results are an application of an exact lattice operator duality recently
developed in Ref.~\cite{Chen2022}, which extended the mapping of Ref.~\cite{Chen2018},
from the underlying local spin degrees of freedom onto non-local fermion (spinon)
and hard-core boson (vison) degrees of freedom.
For related ideas and developments see also Refs.~\cite{Pozo2021,Rao2021,Kapustin2020}.
Our methods could also be useful for investigating other experimentally relevant
observables, such as the static and dynamic spin correlation
functions~\cite{Baskaran2007,Tikhonov2011,Mandal2011,Knolle2015,Song2016}.

Our paper is organized as follows:
In \Secref{sec:model}, we introduce and reviw the model of interest, which is the
Kitaev model perturbed by a Zeeman field. We then discuss how the Zeeman couplings 
could induce vison hoppings in \Secref{sec:vison-hop}, and we set up the general
formulas that define the vison hopping amplitudes and the
phases associated with vison hopping around small closed loops of the lattice.
In \Secref{sec:duality}, we review the recently developed exact duality mapping~\cite{Chen2022}, and discuss how to explicitly compute the vison hopping amplitudes and phases defined in \Secref{sec:vison-hop}.
\Secref{sec:results} contains the main results of this study.
\Secref{sec:FM-AFM-hopping} presents the explicit numerical results of the vison hopping amplitudes and phases for both
FM and AFM coupled models. In \Secref{sec:vison-Chern} we show that for the AFM case the vison bands are topologically non-trivial and carry a finite Berry curvature and non-zero Chern numbers. In \Secref{sec:commuting-projector}
we provide additional evidence for our conclusions, by showing  that the vison phases around loops agree with those of certain commuting projector Hamiltonians
where calculations can be performed fully analytically.
Finally, in \Secref{sec:discuss}, we summarize our findings and make some suggestions for guiding experimental efforts to further determine the sign of $K$ in $\alpha$-RuCl$_3$.
\Appref{sec:PBC} discusses dependence on boundary conditions for both FM and AFM models, and \Appref{sec:finite-size} discusses details of finite-size effects and the infinite size extrapolation of our results.
% 
% ------------------------------------------------------------------------------
% FIGURE
% ------------------------------------------------------------------------------
% ------------------------------------------------------------------------------

\section{Kitaev model Perturbed by a Zeeman field} \label{sec:model}

The model of interest is the Kitaev honeycomb model with a Haldane
mass term~\cite{Kitaev2006}:
\begin{align} \label{eq:H_K}
& H = H_K - \kappa \sum_{i,j,k} X_i Y_j Z_k, \\ \nonumber
& H_K = K \left( \sum_{x \text{-links}} X_i X_j + \sum_{y \text{-links}} Y_i Y_j
+ \sum_{z \text{-links}} Z_i Z_j \right).
\end{align}
Here $X_i$, $Y_j$, $Z_k$ are the Pauli matrices of spins residing in the sites of the
honeycomb lattice (see \Fref{fig:schematic}(a)).
The above Hamiltonian is exactly solvable, and features a dispersive band of
itinerant Majorana fermions and a gapped flat band of visons with an energy
$E_0 \approx 0.15 |K|$ for $|\kappa| \ll |K|$~\cite{Kitaev2006}.
The term $\kappa$ induces a Haldane mass on the Majorana fermions that would otherwise
have a gapless dispersion (see Ref.~\cite{Kitaev2006} for the choice
of $i,j,k$ in the summation).
Therefore this term is needed in order to have a  fully gapped topologically
ordered state and exponentially localized Majorana zero modes carried by the vison.
Throughout the paper we will keep $\kappa$ as an \emph{independent} parameter of the model,
but we will view it implicitly as a term that is perturbatively generated by
a physical Zeeman field, whose leading form is
$\kappa \approx h_x h_yh_z/K^2$~\cite{Kitaev2006}, and in particular in all of our
subsequent discussion it is understood that we fix $\sgn(\kappa)=\sgn(h_xh_zh_y)$.
It should be noted that in the presence of other types of perturbations, the
induced Haldane mass term can scale linearly with $h$ \cite{Song2016}, therefore
a finite $\kappa$ term is physically relevant in those cases as well.

In addition to the above, we include the following explicit Zeeman coupling to
$H$ in \Eqref{eq:H_K}:
\begin{equation} \label{eq:H_Zeeman}
\Delta H = -\sum_j (h_x X_j + h_y Y_j + h_z Z_j).
\end{equation}
The above term is produced by an external magnetic field.
Crucially, this term induces not only the aforementioned Haldane mass term
(provided that each component of the magnetic field is non-zero),
but also the motion and pair creation/annihilation of the vison particles,
destroying the exact solvability of the model. We will therefore treat
this term as a perturbation.
Our goal is to compute the perturbative hoppings and band
dispersions that that this term induces on the visons, and to compute the real space phases that result from such
vison motion when it is transported around a unit cell of its triangular lattice. While the magnitude of these hoppings will depend on the detailed form of the perturbation that induces the vison hopping,
we wish to emphasize that
this phase is independent on the specific detailed form of the operator
that induces the vison motion, as long as it respects translational symmetry,
since this phase is a universal property of the topologically ordered state enriched
by translational symmetry, as recently argued in Ref.~\cite{Chen2022}.
Therefore, as we will see, this phase is in exact agreement
with the phase that was computed in Ref.~\cite{Chen2022} with a perturbation
different from the Zeeman field.

% ------------------------------------------------------------------------------
% FIGURE
% ------------------------------------------------------------------------------
\begin{figure}
\centering
\includegraphics[width=0.49 \textwidth]{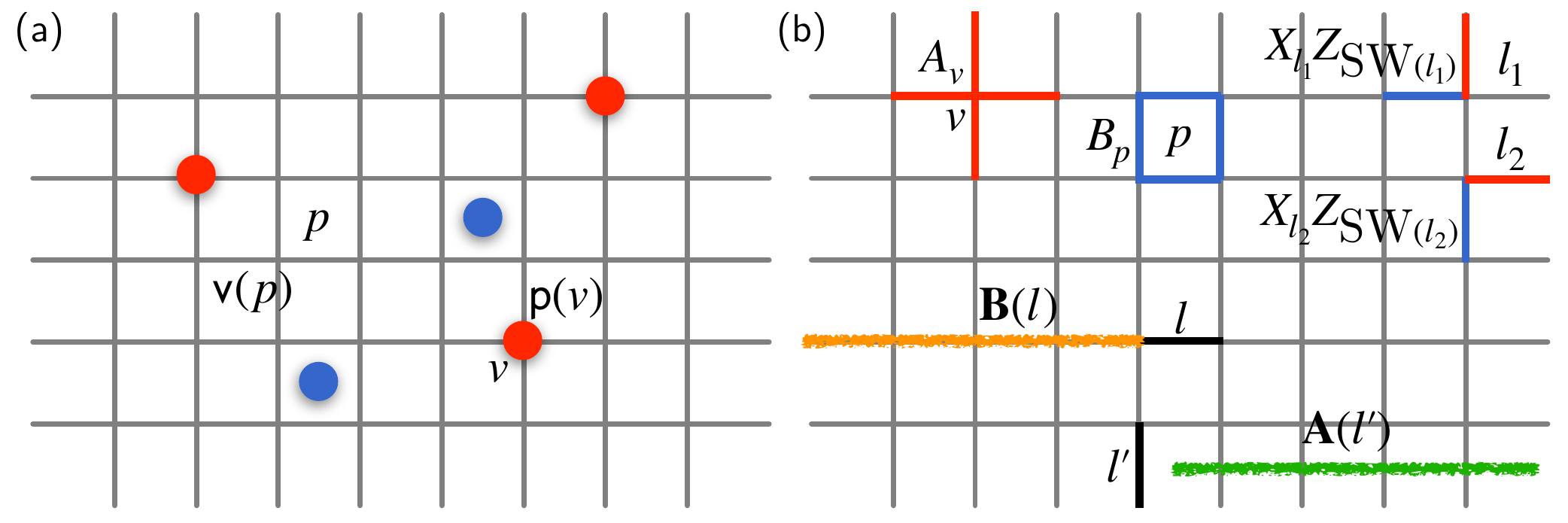}
\caption{
(a) Schematic of a real space configuration of the $e$ (red dots)
and $\varepsilon$ (blue dots) particles.
At each vertex $v$, there is a boson ($e$) mode with creation/annihilation
operators $b_v^\dagger/b_v$.
Within each plaquette $p$, there is a complex fermion ($\varepsilon$) mode
with creation/annihilation operator $c_p^\dagger/c_p$, which is equivalent
to two Majorana modes $\gamma_p$ and $\gamma'_p$.
(b) Spin operators and paths involved in the duality mapping.
The Pauli $X$ ($Z$) matrices at a link is represented by a red (blue) 
colored bond covering the link.
For a vertical/horizontal link $l_{1/2}$, $\SW(l_{1/2})$ is the horizontal/vertical
link to its southwest which connects to it.
The path $\mathbf{B}(l)$ (orange thick line) contains all the vertices to the
left of a horizontal link $l$ and the dual lattice path $\mathbf{A}(l')$
(green thick line) contains all the plaquettes to the right of a vertical link $l'$.
}\label{fig:schematic-square}
\end{figure}
% ------------------------------------------------------------------------------

\section{Methods}
\subsection{Vison hoppings in the honeycomb lattice} \label{sec:vison-hop}
The visons can be viewed as being located at the center of each
hexagon.
The vison parity operator at a hexagon/plaquette $p$, $W_p$
equals $-1\ (1)$ when a vison is present (absent) in a plaquette, and it
is a product of the Pauli matrices of the $6$ spins surrounding $p$~\cite{Kitaev2006}.
For example, for the hexagon $d$ in \Fref{fig:schematic}(a), we have
\begin{equation} \label{eq:W_p}
W_d = X_4 Y_6 Z_7 X_8 Y_1 Z_2.
\end{equation}
For $H$ in \Eqref{eq:H_K}, the vison parity $W_p$ is a constant of motion.
However, its value ($\pm1$) can be flipped by local spin operators.
For example, for an $\alpha$-link ($\alpha = X, Y, Z$) between vertices $i$ and $j$,
$\alpha_i$ and $\alpha_j$ anti-commute with vison parities at both
plaquettes sandwiching the link $(i,j)$, and commute with the vison parities
at other hexagons.
Therefore, $\alpha_{i/j}$ can induce vison hoppings across the link $(i,j)$.
The Zeeman coupling $\Delta H$ in \Eqref{eq:H_Zeeman} involves a sum of
all the local spin operators, as a result, it can also induce
vison hopping across each link.
Below we will illustrate how to extract the vison hopping amplitudes, induced by
$\Delta H$, from an example.

Let us consider an eigenstate of $H$ from \Eqref{eq:H_K} consisting of two far
distant visons. We are interested in the ``single-particle'' properties of the
visons, therefore, we will take one of these visons simply as an auxiliary vison
that is held immobile while the other is allowed to move (this can be accomplised
by only acting with the perturbation $\Delta H$ from \Eqref{eq:H_Zeeman} within
the region containing the mobile vison of interest).
We place the immobile vison at a hexagon at $\bR_0$, and consider fluctuations of
the mobile vison within the hexagon $a$, $b$, $c$ or $d$ in \Fref{fig:schematic}(a).
The \emph{lowest} energy eigenstate of $H$ for each two-vison configuration
is denoted as $| \Phi(\bR_0, \bR_{l}) \rangle$, with $l \teq a, b, c, d$.
In the limit in which both visons are much farther than the typical localization
length of their Majorana zero modes, these four states will be degenerate.
Moreover, when $\kappa$ is finite, there will be a gap to all excitations and
therefore only these configurations that differ by changing the position of the
vison will be connected by the perturbation $\Delta H$.
Under these conditions one can then conceptualize these
processes as coherent hopping of the vison particles.
Then the leading in $\Delta H$ perturbative hopping amplitude of
the vison from $d$ to $b$ (due to $\Delta H$) is given by
\begin{equation} \label{eq:t_eff}
t_{b,d} \equiv
\langle \Phi(\bR_0, \bR_{b}) | \Delta H| \Phi(\bR_0, \bR_{d}) \rangle.  
\end{equation}
From the definition of vison parity operator discussed above,
one can see that only $Y_2$ and $Y_4$ anticommute with $W_d$ and $W_b$ while
commuting with vison parities at all other hexagons. Therefore, 
\begin{align} \label{eq:t_bd}
t_{b,d}
= -h_y \langle \Phi(\bR_0, \bR_{b}) | Y_2 + Y_4 
| \Phi(\bR_0, \bR_{d}) \rangle.
\end{align}
Similarly, it is straightforward to show the following expressions for the
other hoppings:
\begin{subequations} \label{eq:t-honeycomb}
\begin{align}
& t_{a,b} = -h_z
\langle \Phi(\bR_0, \bR_{a}) |Z_3 + Z_2| \Phi(\bR_0, \bR_{b}) \rangle, \\
& t_{b,c} = -h_x
\langle \Phi(\bR_0, \bR_{b}) |X_4 + X_5| \Phi(\bR_0, \bR_{c}) \rangle.
\end{align}
\end{subequations}
Once the effective vison hopping for each link is obtained, one can extract the vison
Berry phase acquired in a closed hopping associated with each closed path.
For example, the phase acquired by hopping in the triangle
$d \rightarrow b \rightarrow a \rightarrow d$ is 
defined as:
\begin{equation} \label{eq:phiut}
\phi_{a b d} = \Im \ln( t_{d,a} t_{a,b} t_{b,d}),
\end{equation}
and the phase acquired by hopping on the unit cell of translations
$d \rightarrow c \rightarrow b \rightarrow a \rightarrow d$ is:
\begin{equation} \label{eq:phi_p}
\phi_{a b c d} = \Im \ln ( t_{d,a} t_{a,b} t_{b,c} t_{c,d} ).
\end{equation}
If the phase $\phi_{a b c d}$ aquired around a unit cell of the
microscopic translational symmetries is not $0 \pmod{2\pi}$,
then the translations have a projective implementation on the vison
particle~\cite{Wen2002,Wen2002a}.
As discussed in Refs.~\cite{Essin2013,Rao2021,Chen2022} for topologically ordered
states with deconfined $Z_2$ gauge fields, this phase is expected to be either $0$
or $\pi$ and its value is a robust topological characteristic of the phase as
long as the the microscopic translational symmetry of the model is enforced.
Notice, however, that these considerations do not apply to the phase $\phi_{abd}$
acquired around a single triangle, because this motion cannot be generated by
the elements of the microscopic translation group, and therefore this phase can
be sensitive to perturbations when only the translational symmetry is enforced.

The above formulae then define the mathematical problem at hand.
The problem of computing these overlaps has been recently addressed in
Ref.~\cite{Joy2021} employing the parton representation of spins.
In the next section, we will however employ a different method that relies in an
exact lattice duality developed in Ref.~\cite{Chen2022} building on the previous
work of Refs.~\cite{Chen2018,Rao2021}.

\subsection{Vison hoppings from the duality mapping} \label{sec:duality}
To use the duality mapping of Ref.~\cite{Chen2022},
it is convenient to change the usual basis of the Kitaev model by performing a unitary transformation $U$.
The transformation $U$ \emph{only} affect the spins from $A$-sublattice sites
(see \Fref{fig:schematic}(a)):
\begin{equation} \label{eq:U-transform}
X_j \leftrightarrow Z_j, \ Y_j \rightarrow -Y_j, \quad \forall j \in A.
\end{equation}
The spin Hamiltonians in \Eqref{eq:H_K} and \Eqref{eq:H_Zeeman} will be transformed 
accordingly, i.e.,
$H_K \rightarrow U H_K U^{-1} =  \tilde{H}_K$,
$\Delta H \rightarrow U \Delta H U^{-1} = \Delta \tilde{H}$:
\begin{align} \label{eq:tilde-H}
\tilde{H}_K = &
K \left(
\sum_{\substack{x\text{-link}, \\ i \in A}} Z_i X_j
- \sum_{ \substack{y\text{-link}, \\ i \in A} } Y_i Y_j
+ \sum_{ \substack{z\text{-link}, \\ i \in A} } X_i Z_j
\right), \nonumber \\
\Delta \tilde{H} = &
- \sum_{i \in A} ( h_z X_i - h_y Y_i + h_x Z_i) \nonumber \\
& - \sum_{j \in B} (h_x X_j + h_y Y_j + h_z Z_j),
\end{align}
and similarly for the $\kappa$ term. In this new basis the vison parity reads as:
\begin{align} \label{eq:W_p-transformed}
W_d \rightarrow \tilde{W}_d & = Z_4 Y_6 X_7 X_8 (-Y_1) Z_2    \nonumber \\
& = X_6 X_7 X_8 X_1 Z_1 Z_2 Z_4 Z_6.
\end{align}
And therefore, by identifying the \emph{sites} of the honeycomb lattice with the
\emph{links} of a square lattice as depicted in \Fref{fig:schematic}(b),
this operator can be
viewed as a product of the star ($A_d$) and plaquette ($B_{\pp(d)}$)
operators of the standard toric code model~\cite{Kitaev2003}
(see \Fref{fig:schematic-square}(b)), as discussed in
Refs.~\cite{Chen2018,Pozo2021,Rao2021,Chen2022}.

We will review here the exact duality mapping introduced in Ref.~\cite{Chen2022}
for the case of an \emph{infinite} system, and refer the reader to
Ref.~\cite{Chen2022} for details of the mapping on a finite open and periodic
lattices.
The lattice duality allows to map the tensor product Hilbert space of spins onto a
tensor product Hilbert space of the ``visons'' and ``spinons''.
The vison is a spinless hard-core boson which can viewed as being located at the 
vertices of the square lattice (see \Fref{fig:schematic-square}(a)),
analogous to the ``$e$-particles'' of the toric code.
Therefore we assign vison hard-core boson creation/annihilation operators,
$b_v^\dagger$, $b_v$,
with every vertex ``$v$''. The vison parity operator is then mapped as follows:
\begin{equation} \label{eq:W_v-dual}
\tilde{W}_v \leftrightarrow e^{ i \pi b_v^\dagger b_v }.
\end{equation}

On the other hand the spinon degrees of freedom correspond to those of a single
spinless complex fermion mode per unit cell (which can be viewed as a descending
of the ``$\varepsilon$-particle'' of the toric 
code~\cite{Chen2018,Pozo2021,Rao2021,Chen2022}).
Therefore we introduce two spinon Majorana fermion modes, $\gamma_p, \gamma'_p$ for
every plaquette ``$p$'' of the lattice as depicted in \Fref{fig:schematic-square}(a).
The fermion parity maps onto the following plaquette operator:
\begin{equation}
B_{\pp(d)}=Z_1 Z_2 Z_4 Z_6 \leftrightarrow -i \gamma_{\pp(d)} \gamma'_{\pp(d)}.    
\end{equation}
Here the operators $Z_1 \cdots Z_6$ are those appearing in the usual plaquette
term of the toric code, $\pp(d)$ stands for the plaquette to the \emph{northeast} of
vertex $d$, as depicted in \Fref{fig:schematic}(b).
We can combine the Majorana modes into complex fermion operators as follows:
\begin{align}
c_p = (\gamma_p + i \gamma'_p)/2, \
c_p^\dagger = ( \gamma_p - i \gamma'_p )/2.
\end{align}
Note that $-i \gamma_p \gamma'_p \teq e^{i \pi c_p^\dagger c_p}$.

We will now present the ``dictionary'' that allows to map exactly any local
operator acting the underlying physical spin degrees of freedom onto operators
acting on the dual vison and spinon degrees of freedom, introduced in
Ref.~\cite{Chen2022}.
The mapping of spin operators on a link $l$ is the following:
\begin{enumerate}[label = \roman*).]
\item Vertical $l$:
\begin{subequations} \label{eq:v-l}
\begin{align}
& Z_l \leftrightarrow (b_{v_1} + b_{v_1}^\dagger) (b_{v_2} + b_{v_2}^\dagger)
e^{i \pi \alpha_l}
\\
& X_l Z_{\SW(l)} \leftrightarrow i \gamma_{p_1} \gamma'_{p_2}.
\end{align}
\end{subequations}
\item Horizontal $l$:
\begin{subequations} \label{eq:h-l}
\begin{align}
& Z_l \leftrightarrow (b_{v_1} + b_{v_1}^\dagger) (b_{v_2} + b_{v_2}^\dagger)
\\
& X_l Z_{\SW(l)} \leftrightarrow
e^{i \pi \beta_l}
i \gamma_{p_1} \gamma'_{p_2}.
\end{align}
\end{subequations}
\end{enumerate}
Where:
\begin{align}
\alpha_l = \sum_{p \in \bA(l)} c_p^\dagger c_p, \ 
\beta_l = \sum_{v \in \bB(l)} b_v^\dagger b_v.
\end{align}
Here for a vertical (horizontal) link $l$, $v_1$ and $v_2$ are the two vertices
connected by it, while $p_1$ and $p_2$ are the
two plaquettes at the left and right (top and bottom) sides of it,
$\SW(l)$ is the link to the southwest of $l$ which also connects to it
(see \Fref{fig:schematic-square}(b)).
Also $\bA(l)$ ($\bB(l)$) denotes sets of plaquettes (vertices) that reside on a
string in the dual (direct) lattice and are depicted in
\Fref{fig:schematic-square}(b).
These ``string'' operators are the ones encoding that the vison and the spinon 
have a non-local statistical interactions that makes them behave as mutual semions.
Notice that the spin operators described above form a complete algebraic basis
out of which any other operator can be obtained by taking sums, products and
multiplication by complex numbers, from this basis.

In particular, we can apply the above dictionary to re-write the Hamiltonian from
\Eqref{eq:tilde-H} in terms of vison and spinon degrees of freedom, leading to:
\begin{widetext}
\begin{align} \label{eq:H_dual}
\tilde{H} = &
K \sum_p \left[ e^{i \pi \beta_{\tilde{L}(p,p+\hat{y})}}
( i \gamma_{p+\hat{y}} \gamma'_p + i \gamma_{p+\hat{y}} \gamma'_{p+\hat{x}})
 + i\gamma_p \gamma'_{p+\hat{x}} \right]
\nonumber \\
& - \kappa \sum_p \left[
e^{ i \pi \beta_{\tilde{L}(p,p+\hat{y})} }
( i \gamma_{p+\hat{y}} \gamma_{p-\hat{{x}}})
+ e^{i \pi b_{\vv(p)}^\dagger b_{\vv(p)}} (i \gamma_{p-\hat{x}} \gamma_p )
+ e^{ i \pi \beta_{\tilde{L}(p,p+\hat{y})} }
( i \gamma_p \gamma_{p+\hat{y}} ) \right.
\nonumber \\
& \left. 
+ e^{ i \pi \beta_{\tilde{L}(p-\hat{x},p-\hat{x}+\hat{y})} }
(i \gamma'_{p+\hat{y}} \gamma'_p )
+ i \gamma'_{p+\hat{x}} \gamma'_p
+ e^{ i \pi \beta_{\tilde{L}(p,p-\hat{y})} }
(i \gamma'_{p-\hat{y}} \gamma'_{p+\hat{x}} ) \right].
\end{align}
\end{widetext}
Here $\vv(p)$ denotes the vertex to the \emph{southwest} of plaquette $p$
(see \Fref{fig:schematic-square}(a)),
$\tilde{L}(p,p')$ stands for the link sandwiched by plaquettes $p$ and $p'$.
Notice that the above Hamiltonian explicitily commutes with the vison occupation
of all the vertices in the lattice and the remaining dynamical degrees freedom are
described by a fermion bilinear model, as expected.

On the other hand, one can show that the $\Delta \tilde{H}$ in \Eqref{eq:tilde-H}
can be re-written as:
\begin{widetext}
\begin{align} \label{eq:DeltaH_dual}
\Delta \tilde{H} = & - \sum_v \left[
h_z (b_v + b_v^\dagger)(b_{v+\hat{x}} + b_{v+\hat{x}}^\dagger) 
(1 + i\gamma_{\pp(v)} \gamma'_{\pp(v)+\hat{x}})
+ h_y (b_{v-\hat{x}}+b_{v-\hat{x}}^\dagger)(b_{v+\hat{y}} + b_{v+\hat{y}}^\dagger)
(\gamma_{\pp(v)-\hat{x}} \gamma'_{\pp(v)}) 
e^{ i \pi \alpha_{L(v,v+\hat{y})} }
\right. \nonumber \\
& + h_y e^{ i \pi \beta_{L(v+\hat{y},v+\hat{y}+\hat{x})} }
(b_v + b_v^\dagger)(b_{v+\hat{x}+\hat{y}}+b_{v+\hat{x}+\hat{y}}^\dagger)
(-\gamma_{\pp(v)+\hat{y}} \gamma'_{\pp(v)}) e^{ i \pi \alpha_{L(v,v+\hat{y})} }
\nonumber \\
& \left. + 
h_x \left( e^{ i \pi \beta_{L(v+\hat{y},v+\hat{y}+\hat{x})} }
i\gamma_{\pp(v)+\hat{y}} \gamma'_{\pp(v)} + 1 \right)
(b_v + b_v^\dagger)(b_{v+\hat{y}}+b_{v+\hat{y}}^\dagger)
e^{ i \pi \alpha_{L(v,v+\hat{y})} } \right].
\end{align}
\end{widetext}
Here $L(v,v')$ stands for the link connecting vertices $v$ and $v'$.
As we see, the Zeeman term contains vison hopping and pair creation terms,
creating an impediment to solving it exactly. 

% As shown in \cref{eq:H_dual,eq:DeltaH_dual}, $\Delta\tilde{H}_\text{dual}$
% is able to hop the $e$ particles/visons, as it contains terms of the form:
% $(b_i + b_i^\dagger) (b_j + b_j^\dagger)$.
% For $\tilde{H}_\text{dual}$, since the vison parity at each vertex is a good
% quantum number and the fermionic part consists of only bilinear terms,
% it can be solved exactly within any sector of $e$-particle configuration.
% Considering a sector with $N$ $e$-particles located at
% $\{ v_1, \dots, v_{N} \}$, the problem then reduces to that of
% diagonalizing a Bogoliubov de-Gennes (BdG) Hamiltonian
% $H^\varepsilon( v_1,\dots,v_{N} )$ with $\pi$ fluxes located at
% $\{ v_1,\dots,v_{N} \}$.
% $H^\varepsilon( v_1,\dots,v_{N} )$ is obtained from \Eqref{eq:H_dual}
% by replacing 
% \begin{equation}
% b_{v}^\dagger b_{v} \rightarrow
% \begin{cases}
% 1, & \text{if}\ v \in \{ v_1, \dots, v_{N_e} \} \\
% 0, & \text{others}.
% \end{cases}
% \end{equation}
% An eigenstates of $\tilde{H}_\text{dual}$ has the following form:
% \begin{equation}
% b_{v_1}^\dagger \dots b_{v_{N}}^\dagger | 0 \rangle \otimes
% | \Psi^\varepsilon(v_1,\dots,v_{N}) \rangle,
% \end{equation}
% where $|\Psi^\varepsilon(v_1,\dots,v_{N}) \rangle$ is an eigenstate of
% $H^\varepsilon(v_1,\dots,v_{N})$.

Therefore our strategy is to treat $\Delta \tilde{H}$ as a perturbation acting on the
exact eigenstates of \Eqref{eq:H_dual}.
Let us discuss how to uniquely label these states. An eigenstate of \Eqref{eq:H_dual} with $N$ visons placed in the vertices $\{ v_1,....,v_N \}$, can be written as a
tensor product of vison and spinon degrees of freedom as follows:

\begin{equation}
b_{v_1}^\dagger \dots b_{v_{N}}^\dagger | 0 \rangle \otimes
| \Psi^\varepsilon(\bR_1, \dots, \bR_{N}) \rangle,
\end{equation}
where $\bR_i$ denotes the position of vertex $v_i$.
The first term specifies the vison locations and the second term is the fermionic
spinon wavefunction resulting from diagonalizing the effective BdG Hamiltonian
$H^\varepsilon(\bR_1, \dots, \bR_{N})$ associated with \Eqref{eq:H_dual}.
Notice that this Hamiltonian has already a unique specified gauge choice for the
vector potential that captures the long-range statistical interaction after the
vison occupations are given and viewed as constants.
We caution that the tensor product structure above is in a dual Hilbert space
and does not have a simple relation to the tensor structure of the underlying physical 
spin degrees of freedom (see Ref.~\cite{Chen2022} for further discussions).

Let us now specialize to the case of two visons to compute the matrix elements
described in \cref{eq:t_eff,eq:t_bd,eq:t-honeycomb}.
These vison hopping elements can then systematically computed. For example the
hopping between vertices $a$ and $b$ (see \Fref{fig:schematic}(b))
can be obtained as follows:
\begin{align} \label{eq:t_ab-dual}
t_{a,b} = & -h_z
\langle 0| b_0 b_a \otimes \langle \Psi^\varepsilon_0(\bR_a, \bR_0) |
(b_a + b_a^\dagger)(b_b + b_b^\dagger) \nonumber \\
& (1+i\gamma_{\pp(a)} \gamma'_{\pp(b)}) \ 
b_b^\dagger b_0^\dagger | 0 \rangle
\otimes | \Psi^\varepsilon_0 (\bR_b, \bR_0) \rangle \nonumber \\
= & -h_z \langle \Psi^\varepsilon_0(\bR_a, \bR_0) |
( 1 + i \gamma_{\pp(a)} \gamma'_{\pp(b)} )
| \Psi^\varepsilon_0 (\bR_b, \bR_0) \rangle.
\end{align}
Here $| \Psi^\varepsilon_0( \bR_j, \bR_0) \rangle$ is the \emph{lowest} energy
eigenstate of $H^\varepsilon(\bR_j, \bR_0)$.
Similarly, the hopping between vertices $a$ and $d$ reads:
\begin{align} \label{eq:t_da-dual}
t_{d,a} = & -h_x
\langle \Psi^\varepsilon(\bR_d,\bR_0)| (1 + i \gamma_{\pp(a)} \gamma'_{\pp(d)})
e^{i \pi \alpha_{L(a,d)} } \nonumber \\
& | \Psi^\varepsilon(\bR_a,\bR_0) \rangle.
\end{align}
Here $e^{i \alpha_{L(a,d)}}$ reflects the statistical interaction between
visons and spinons, and we have assumed that the vertex $0$ is in the same row as
vertex $d$.

We would like to briefly comment on the duality mapping for the case of a
periodic system, which is the geometry that we use for the numerical
implementation to be described in the next section.
There are two global constraints on a torus:
\begin{align}
\prod_v \tilde{W}_v = 1, \ \prod_p B_p = 1.
\end{align}
Therefore only states with an \emph{even} number of visons and spinons are physical in
the torus.
Moreover, there are additionally two global Wilson loop degrees of freedom
associated with the non-contractible loops around the torus.
% \begin{equation}
% \mathcal{H}^\text{even}_e \otimes \mathcal{H}^\text{even}_\varepsilon \otimes
% \mathcal{H}_{\text{WLSs}}.
% \end{equation}
% It can be shown that the $\tilde{H}_\text{dual}$ now conserves both $e$-particle
% parity at each vertex and the spin-$Z$ of WLSs: $(z_1,z_2)$.
% Similar to the infinite lattice case,
% in a sector with $e$ particles located at $\{ v_1, \dots, v_{N} \}$ and 
% WLSs being $| z_1,z_2 \rangle$, the $\tilde{H}_\text{dual}$
% reduces to a BdG Hailtonian
% $H^\varepsilon(v_1, \dots, v_{N}; z_1, z_2)$, which contains $\pi$ fluxes
% at $\{ v_1, \dots, v_{N} \}$.
% It can be shown that $z_1$ and $z_2$ will determine the fermmion boundary
% conditions along $x$ and $y$ directions respectively, e.g.,
% periodic/anti-periodic boundary condition along $x$-direction when $z_1 = \pm 1$. 
% When WLSs are $(z_1,z_2)$ and two $e$ particles are located at $\bR_j$ and
% $\bR_0$, one needes to obtain the lowest energy \emph{even}-parity eigenstate of
% $H^\varepsilon(\bR_j,\bR_0;z_1,z_2)$:
For the exactly solvable model where the visons are static, the topological degrees
of freedom associated with the Wilson loops can be taken to be two quantum numbers
with values $z_{1,2} = \pm 1$ that specify whether the spinons have periodic or
anti-periodic boundary conditions along the $x$- or $y$-directions of the torus
respectively.
Therefore the fermionic BdG eigenstates are labeled as:
$|\Psi^\varepsilon_0 (\bR_j,\bR_0;z_1,z_2) \rangle$.
The vison hoppings can then be calculated in the same way as shown in
\cref{eq:t_ab-dual,eq:t_da-dual}.
More details on the duality mapping on a torus can be found in Ref.~\cite{Chen2022}.

In the next section, we will discuss the results with $(z_1,z_2) \teq (-1,-1)$,
i.e., anti-periodic boundary conditions (APBC) for $\varepsilon$ particles.
The results with periodic boundary condition (PBC) ($(z_1,z_2) = (1,1)$)
are presented in \Appref{sec:PBC}.
% 
% ------------------------------------------------------------------------------
% FIGURE
% ------------------------------------------------------------------------------
\begin{figure}[t]
\centering
\includegraphics[width=0.49 \textwidth]{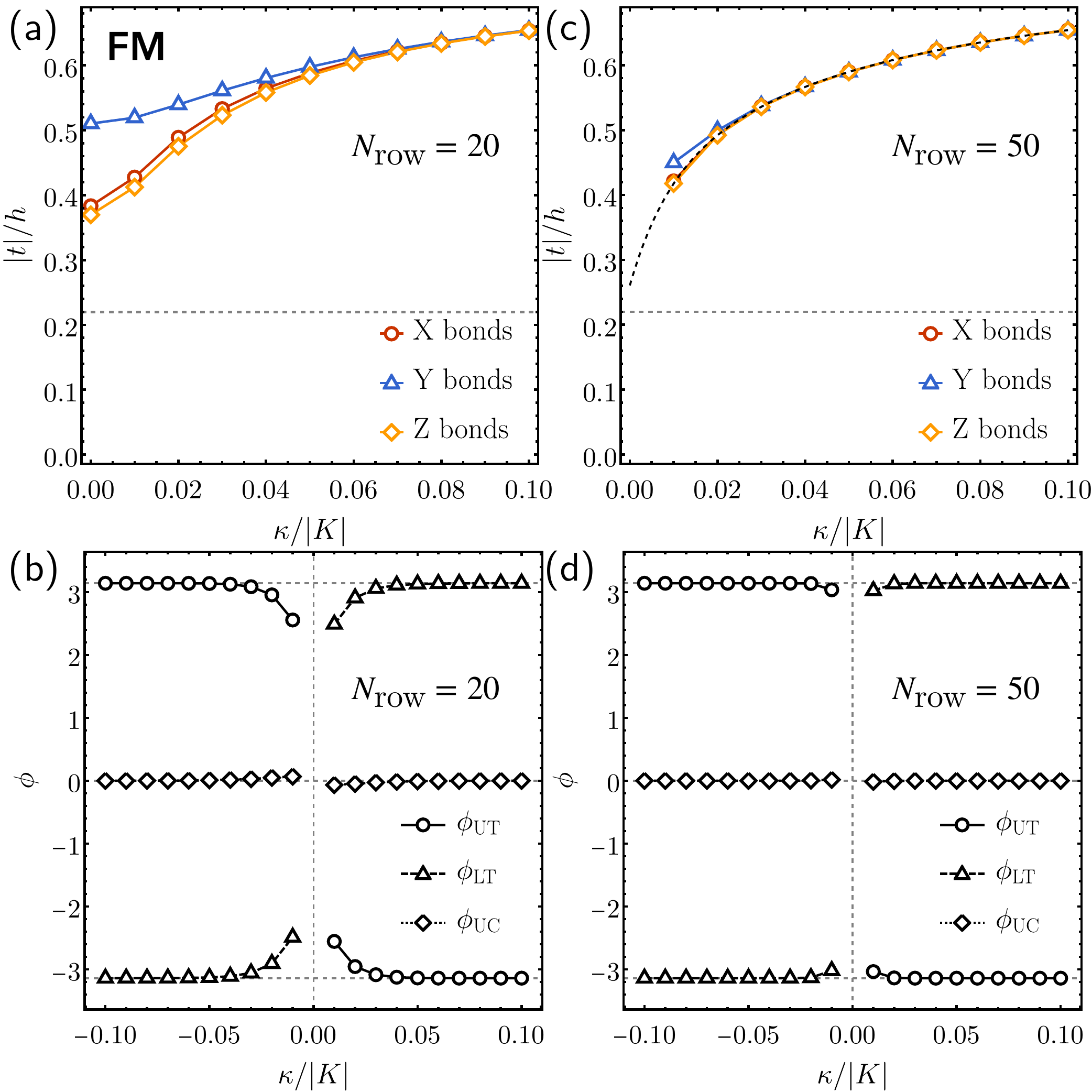}
\caption{
The vison hopping amplitude and Berry phases with FM coupling.
(a) Hopping amplitudes with system size $N_{\text{row}} \teq 20$.
(b) The Berry phases associated with two minimal triangles and the plaquette
path.
(c-d) Same quantities for $N_\text{row} \teq 50$. The magnitude of the hopping is independent of  $\sgn(\kappa)=\sgn(h_xh_zh_y)$, therefore
we only present results for positive $\kappa$.
The amplitudes have been normalized for a Zeeman field
 along the $[111]$ direction: $h_x \teq h_y \teq h_z \teq h/\sqrt{3}$, but the hoppings for other directions of the Zeeman field can be easily estimated from the above plots by re-scaling using Eqs.~\eqref{eq:t_bd}-\eqref{eq:phi_p}. 
}\label{fig:FM_vary_kappa}
\end{figure}
% ------------------------------------------------------------------------------
% 
% ------------------------------------------------------------------------------
% FIGURE
% ------------------------------------------------------------------------------
\begin{figure}
\centering
\includegraphics[width=0.49 \textwidth]{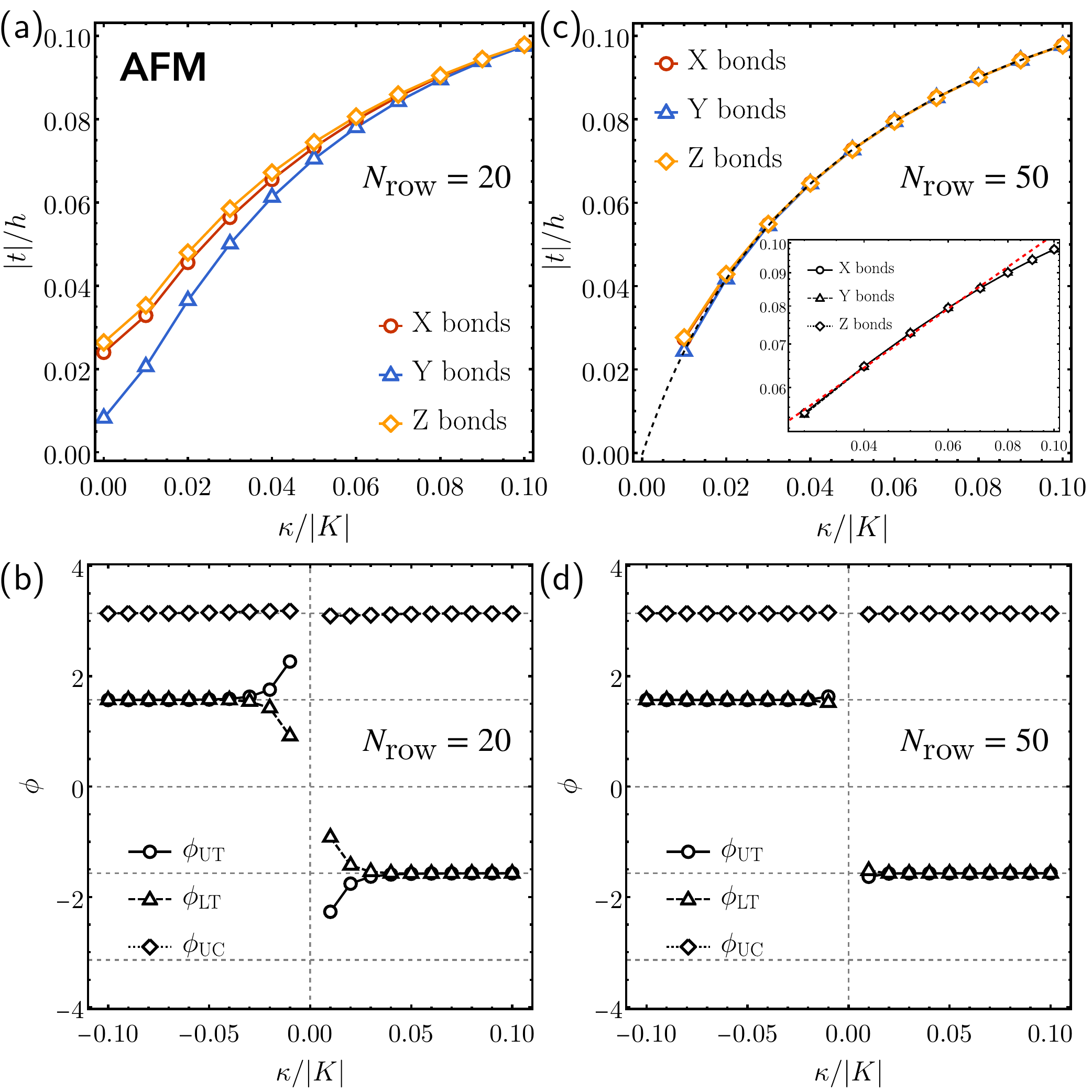}
\caption{
The vison hopping amplitude and Berry phases with AFM coupling.
(a) The hopping amplitude with system size $N_{\text{row}} \teq 20$.
(b) The Berry phases associated with two minimal triangles and the plaquette.
(c-d) Same quantities for $N_\text{row} \teq 50$.
The inset of (c) is a log-log plot of the same data, which suggests
$|t_\alpha| \approx 0.55 |h_\alpha| \ |\kappa/K|^{\nu}$ with
$\nu \approx 0.5$, when the vison hops across an $\alpha$-bond $\alpha \in \{X,Y,Z\}$
(see Fig.~\ref{fig:schematic}). Notice that the vison phases on the triangles depend
on $\sgn(\kappa)=\sgn(h_xh_zh_y)$. We use same the conventions for phases and for
normalizing hoppings as in Fig.~\ref{fig:FM_vary_kappa}.}\label{fig:AFM_vary_kappa}
\end{figure}
% ------------------------------------------------------------------------------
% 
\section{Vison hoppings and Berry phases} \label{sec:results}
\subsection{Results for FM and AFM couplings} \label{sec:FM-AFM-hopping}
As the expressions in \cref{eq:t_ab-dual,eq:t_da-dual} illustrate,
the calculation of the vison hopping amplitudes and Berry phases has been
reduced to a problem of computing matrix elements of operators between free fermion BCS
ground states.
The mathematical details on how to compute these matrix elements have been discussed
in Refs.~\cite{Robledo2009,Chen2022}, and we will here present results following the
same approach of Ref.~\cite{Chen2022}. All the numerical results that we will show have been done for the square torus:
$N_\text{row} \teq N_{\text{col}}$.
For the plots that we will present we have taken the Zeeman field along the $[111]$
direction, so that $h_x \teq h_y \teq h_z \teq h/\sqrt{3}$. Nevertheless, the vison
hoppings for other directions of the Zeeman field can be easily estimated from
the \cref{fig:FM_vary_kappa,fig:AFM_vary_kappa} via a re-scaling using
\cref{eq:t_bd,eq:t-honeycomb,eq:phiut,eq:phi_p} and keeping track of the 
corresponding change of $\kappa \approx h_xh_yh_z/K^2$.

\Fref{fig:FM_vary_kappa} shows the results for the FM Kitaev model at two
different system sizes:
$N_\text{row} \teq 20$ in (a-b), $N_\text{row} \teq 50$ in (c-d).
The hoppings are labeled according to the type of bonds crossed by the vison,
e.g., $t_{a,b} \in Z$-bonds, $t_{b,d} \in Y$-bonds, $t_{d,a} \in X$-bonds
(see \Fref{fig:schematic}(a)).
Our results are consistent with the expected symmetry of the honeycomb lattice
according to which the magnitudes of the hopping amplitude on all bonds
are the same in the thermodynamic limit ($N_\text{row} \rightarrow \infty$),
however the convergence degrades and becomes slower as the fermion gap vanishes for
$\kappa \rightarrow 0$, as evidenced by contrasting the behavior of
$N_\text{row} \teq 20$ with $N_\text{row} \teq 50$ in \Fref{fig:FM_vary_kappa},
and as further discussed in \Appref{sec:finite-size}~\footnote{
We only accessed $\kappa \teq 0$ for
$N_\text{row} \lesssim 20$ due to the slow convergence. This is why this point is missing for
$N_\text{row} \teq 50$ in \Frefs{fig:FM_vary_kappa}(c-d)}. Moreover, we have observed that for strict $\kappa \teq 0$ the vison hoppings are sensitive to the choice of fermion boundary conditions 
(Wilson loop sectors $(z_1,z_2)$), further indicating that the vison hopping may not be
well defined in the $\kappa \teq 0$ case (see \Appref{sec:PBC} for more details). The horizontal dashed lines in \Fref{fig:FM_vary_kappa}(a) and (c)
indicate the result of Ref.~\cite{Joy2021} for the magnitude of the hopping,
which performed all calculations strictly at $\kappa \teq 0$. We see that our extrapolation to $\kappa \rightarrow 0$ at $N_\text{row} \teq 50$
(dashed curve in \Fref{fig:FM_vary_kappa}(c)) 
is in agreement with Ref.~\cite{Joy2021}, and is given by $|t_\alpha| \sim 0.38 |h_\alpha|$, when the vison hops across an $\alpha$-bond $\alpha \in \{X,Y,Z\}$ (see Fig.~\ref{fig:schematic}).

The phases for the FM model are shown in \Fref{fig:FM_vary_kappa}(b) and (d), for a vison hopping around an upper triangle $\phiut$
($d \rightarrow b \rightarrow a \rightarrow d$ in \Fref{fig:schematic}),
a lower triangle $\philt$ ($d \rightarrow c \rightarrow b \rightarrow d$
in \Fref{fig:schematic}), and a Bravais unit cell $\phiuc$
($d \rightarrow c \rightarrow b \rightarrow a \rightarrow d$ in
\Fref{fig:schematic}). We see clear evidence that for $\kappa \neq 0$ these phases approach the following values in the thermodynamic limit:
\begin{subequations}
\begin{align}
\philt &= -\phiut=\pi \sgn(\kappa)=\pi, \ {\rm(FM)} \label{eq:phitFM} \\
\phiuc &= 0, \ {\rm(FM)} \label{eq:phipFM}
\end{align}
\end{subequations}
where in the last equality of \Eqref{eq:phitFM} we have used the fact that
phases are defined modulo $2\pi$. The above is one of our central findings:
the vison in the FM model acquires zero phase around a unit cell of the
Bravais lattice, and thus translations act in a non-projective fashion.

The results for the AFM coupling are shown in \Fref{fig:AFM_vary_kappa}.
Interestingly, we find that the hopping amplitude approaches $0$ as
$\kappa \rightarrow 0$ in the thermodynamic limit, in agreement with
Ref.~\cite{Joy2021}.
Even more remarkably, for non-zero $\kappa$, our results are approaching
the following values of the 
phases around triangles and the unit cell in the thermodynamic limit:
\begin{subequations}
\begin{align}
\philt & = \phiut=-\frac{\pi}{2} \sgn(\kappa), \ {\rm(AFM)} \label{eq:phitAFM} \\   
\phiuc & = \pi, \ {\rm(AFM)} \label{eq:phipAFM}
\end{align}
\end{subequations} 
Therefore we see that the vison has acquires phase $\pi$ when moving around a unit
cell and therefore translations need to be implemented projectively.
Similar to the FM case, in the AFM model at $\kappa \teq 0$ we  also observe strong
sensitivity on the spinon boundary conditions and system sizes
(see \Appref{sec:PBC} for more details).
%$\phiut \teq \philt \teq \mp \pi/2$ for positive (negative) $\kappa$, and $\phiuc \teq \pi$. Therefore, in this case the vison would indeed have a projective symmetry impletations of translations with $\pi$ flux per unit cell.

%Similar to the FM coupled case, extrapolating the finite $\kappa$ data to $\kappa \rightarrow 0$ with a positive $h$ (a $\pi$ shift to $\phiut$ and $\philt$ values at negative $\kappa$) also tends to indicate a discontinuity. Direct calculations at $\kappa \teq 0$ are also performed at $N_\text{row} \le 20$, the results also exhibit a strong dependence on the spinon boundary conditions and system sizes (see \Appref{sec:PBC} for more discussions).

% ------------------------------------------------------------------------------
% FIGURE
% ------------------------------------------------------------------------------
\begin{figure}
\centering
\includegraphics[width=0.48 \textwidth]{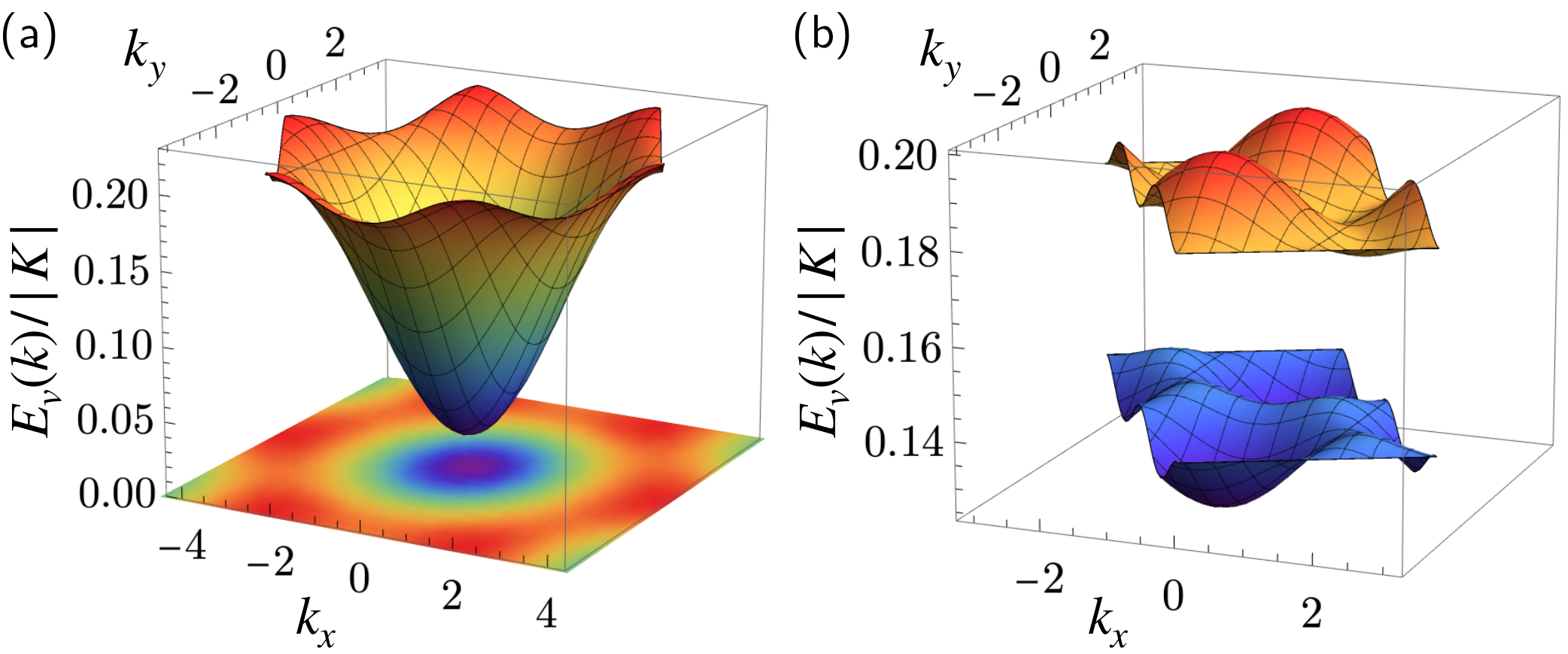}
\caption{
The vison bands for FM (a) and AFM (b) couplings at small Zeeman fields.
The triangular lattice's lattice constant is set to be $a \teq 1$.
(a) The bands for FM coupling with $h = 0.1 |K|$, so
$\kappa/|K| \teq (h/\sqrt{3})^3/|K|^3 \sim 10^{-4}$ according to perturbation
theory~\cite{Kitaev2006}. 
The single vison excitation energy is found to be $E_0 \approx 0.15 |K|$.
(b) The bands for AFM coupling with $\kappa/K = (h/\sqrt{3}K)^3 = 0.01$.
For this $\kappa$ single vison excitation energy is $E_0 \approx 0.16 K$
(obtained for $N_\text{row} = 30$).
Note that for the FM Kitaev model, the vison band has a larger width and
its band minumum is closer to zero.
}\label{fig:vison-bands}
\end{figure}
% ------------------------------------------------------------------------------
% ------------------------------------------------------------------------------
% FIGURE
% ------------------------------------------------------------------------------
\begin{figure}
\centering
\includegraphics[width=0.48 \textwidth]{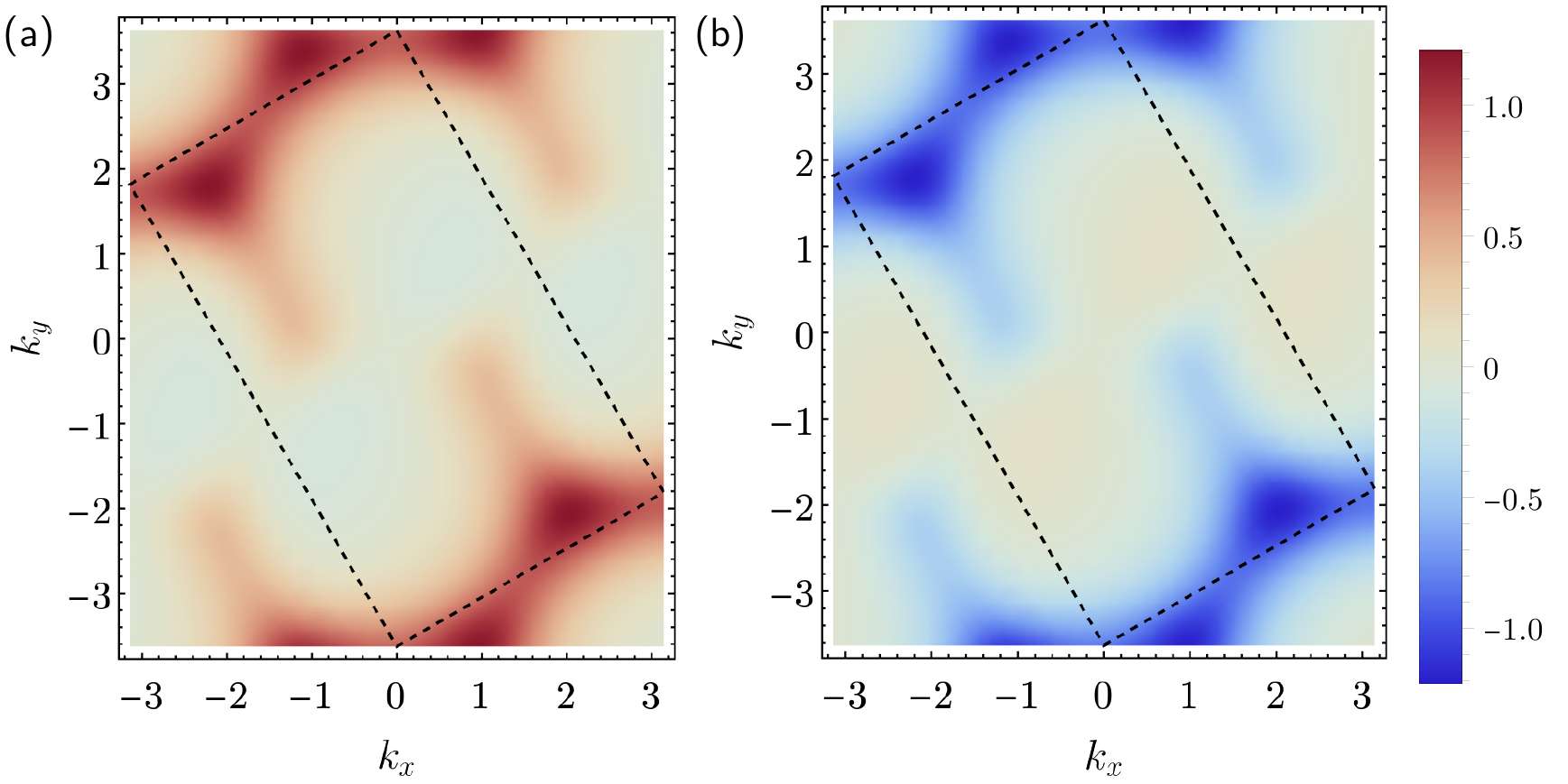}
\caption{
Berry curvature of the upper (a) and lower (b) vison bands in
\Fref{fig:vison-bands}(b). The black dashed rectangle indicates the
reduced Brillouin zone. The Chern number of the upper/lower band is $\pm1$.
}\label{fig:vison-BC}
\end{figure}
% ------------------------------------------------------------------------------
% 
\subsection{Vison Chern bands} \label{sec:vison-Chern}
From the effective vison hoppings computed in the previous section we can construct
the vison band dispersions for both FM and AFM couplings. For FM coupling, the vison does not experience  flux within the unit cell
($\phiuc \teq 0$),
and therefore, the vison band is simply that of a single-site nearest-neighbor tight-binding model in the triangular lattice. As a consequence this band has no Berry curvature. \Fref{fig:vison-bands}(a) shows the vison band dispersion at
$h \teq 0.1 |K|$, $\kappa/|K| \teq (h/\sqrt{3} |K|)^3 \sim 10^{-4}$, which has 
a small gap.

On the other hand, for the AFM model, $\phiut \teq \philt \teq -\pi/2\sgn(\kappa)$ and
$\phiuc \teq \pi$. Therefore the vison unit cell needs to be doubled, giving rise to
two vison bands.
\Fref{fig:vison-bands}(b) shows the vison bands at
$\kappa/K \teq (h/\sqrt{3}K)^3 \teq 0.01$.
We have also calculated the Berry curvature and Chern number for each band.
Interestingly, we found that both bands have a non-trivial topology, the
higher/lower band has a Chern number $C \teq \pm 1$.
The Berry curvatures for the two bands are shown in \Fref{fig:vison-BC}.

\section{Consistency check of vison phases via commuting projector Hamiltonians}
\label{sec:commuting-projector}
In this section we would like to offer some supporting evidence that the phases
that the vison acquires around a unit cell are indeed $0$ and $\pi$ for the FM and
AFM Kitaev models, respectively. Currently we don't know any method allowing us to
compute these phases purely analytically when the fermionic Majorana
spinons are forming a topological 
superconducting state with a non-zero BdG Chern number $C \neq 0$.
Nevertheless, as argued in Ref.~\cite{Rao2021}, when the Majorana spinons have
zero Chern number ($C \teq 0$), the phase that the vison acquires around each unit
cell can indeed be computed analytically thanks to the fact that the state can be
adiabatically deformed into the ground state of a Hamiltonian made out of sums of
commuting projectors (like the toric code), and such adiabatic deformation can be
performed without breaking the lattice translational symmetry of the model.

To exploit this idea, we will add a term to the Hamiltonian that drives a phase
transition into a state with vanishing spinon Chern number ($C \teq 0$) that can
be described with commuting projector Hamiltonians. By following the discussion of
Ref.~\cite{Rao2021}, we will select this commuting projector state such that the
vison still acquires the same phase around a unit cell as in the original state of 
interest with non-zero $C$. By explicitly verifying that the the phase of the vison
around a plaquette indeed does not change across such a phase transition,
we will be able to clearly confirm that the values of the vison phases that we have
computed numerically agree with those of the simpler commuting projector Hamiltonian 
state.

But what are these ideal commuting projector states?
As discussed in Refs.~\cite{Rao2021,Kou2009,Kou2010,Chiu2016}, the gapped paired
BCS states of fermionic spinons with translational symmetry can be classified by the
Chern number
$C \in \mathbb{Z}$ and by four parity indices $\zeta_k = \pm 1$, associated with
the four special high-symmetry points (HSPs) of the Brillouin zone:
$\{ (0,0), (0,\pi), (\pi,0), (\pi,\pi) \}$.
The value $\zeta_k = -1(1)$ is viewed as non-trivial (trivial),
and it indicates that the ground state has an odd (even) number fermions occupying
the special momentum $k$ state
(if such a $k$ state is allowed by the boundary conditions and the size
of the system).
The Chern number and these parity indices are related as follows:
\begin{equation}
(-1)^C = \zeta_{(0,0)} \zeta_{(0,\pi)} \zeta_{(\pi,0)} \zeta_{(\pi,\pi)}.
\end{equation}
The states with commuting projector Hamiltonians of our interest have $C \teq 0$
and either all the parity indices taking the trivial value ($\zeta_k \teq 1$) or
all taking the non-trivial value ($\zeta_k \teq -1$)
\footnote{Those states with only two non-trivial
$\zeta_k \teq -1$ and $C \teq 0$, are ``weak topological superconductors'',
namely stacks of lower dimensional superconductors. These states display 
``weak symmetry breaking''~\cite{Rao2021}, and accordingly, the visons cannot be
transported by local operators to all the nearest neighbor sites of the Bravais
lattice.}. Among these states the one with all the four parity indices trivial
($\zeta_k \teq 1$) and $C \teq 0$ is adiabatically connected to a trivial
``atomic insulator'' vacuum of fermionic spinons, namely the state of spinon
is completely empty, which is essentially a toric code vacuum state,
which we label TC-AI$_0$~\footnote{This state is is also labeled as ``AI$_0$''
in Ref.~\cite{Rao2021} and is labeled as ``$A_+$'' in a
recent general classification of the Abelian $Z_2$ spin liquids
(those with $C = 0$) enriched by lattice translations~\cite{Song2022}}.
As discussed in Refs.~\cite{Rao2021,Chen2022}, this state is adiabatically equivalent
to the ground state of the usual toric code, and the vison in this case is
simply the usual $e$-particle, which clearly acquires $0$ phase when
moving around a unit cell.

% \inti{Original version:
% On the other hand, the states with $\zeta_k = -1$ for all the four HSPs,
% are adiabatically connected to another trivial ``atomic insulator'' state,
% with one $\epsilon$ spinon per unit cell, which is the analogue of a fully occupied
% tight-binding band of spinons, and it is labeled ``AI$_1$'' in Ref.~\cite{Rao2021}.
% The ideal commuting projector Hamiltonians in this case can be taken to be the toric
% code model but with the opposite sign for plaquette coupling, so that every plaquette 
% contains an $m$-particle, relative to the usual toric code (which is  equivalent to a 
% completely filled band of $\epsilon$ fermions in the basis of $\epsilon$ and
% $e$ particles as discussed in Refs.~\cite{Rao2021,Pozo2021, Chen2022}).
% It is straightforward to verify that the vison ($e$-particle) in this state
% will acquire a $\pi$-flux when encircling a unit cell, simply because each unit
% cell contains an $m$-particle, which is a semion relative to the $e$-particles.
% }

On the other hand, the states with $\zeta_k = -1$ for all the four HSPs,
are adiabatically connected to another trivial ``atomic insulator'' state,
with one $\epsilon$ spinon per unit cell, which is the analogue of a fully occupied
toric code state, which we label
TC-AI$_1$~\footnote{This state is labeled as ``AI$_1$'' in Ref.~\cite{Rao2021}
and is labeled as ``$A^\varepsilon$'' in the recent general classification
of the Abelian $Z_2$ spin liquids enriched by lattice translations of
Ref.~\cite{Song2022}.}.
The ideal commuting projector Hamiltonians in this case is the toric
code model but with the opposite sign for both the vertex and plaquette couplings,
so that every plaquette contains an
$\varepsilon$-particle~\cite{Rao2021,Pozo2021, Chen2022}.
It is straightforward to verify that the vison ($e$-particle) in this state
will acquire a $\pi$-flux when encircling a unit cell, simply because each unit
cell contains an $\varepsilon$-particle, which is a semion relative to the
$e$-particles.

\begin{figure}
\centering
\includegraphics[width=0.48\textwidth]{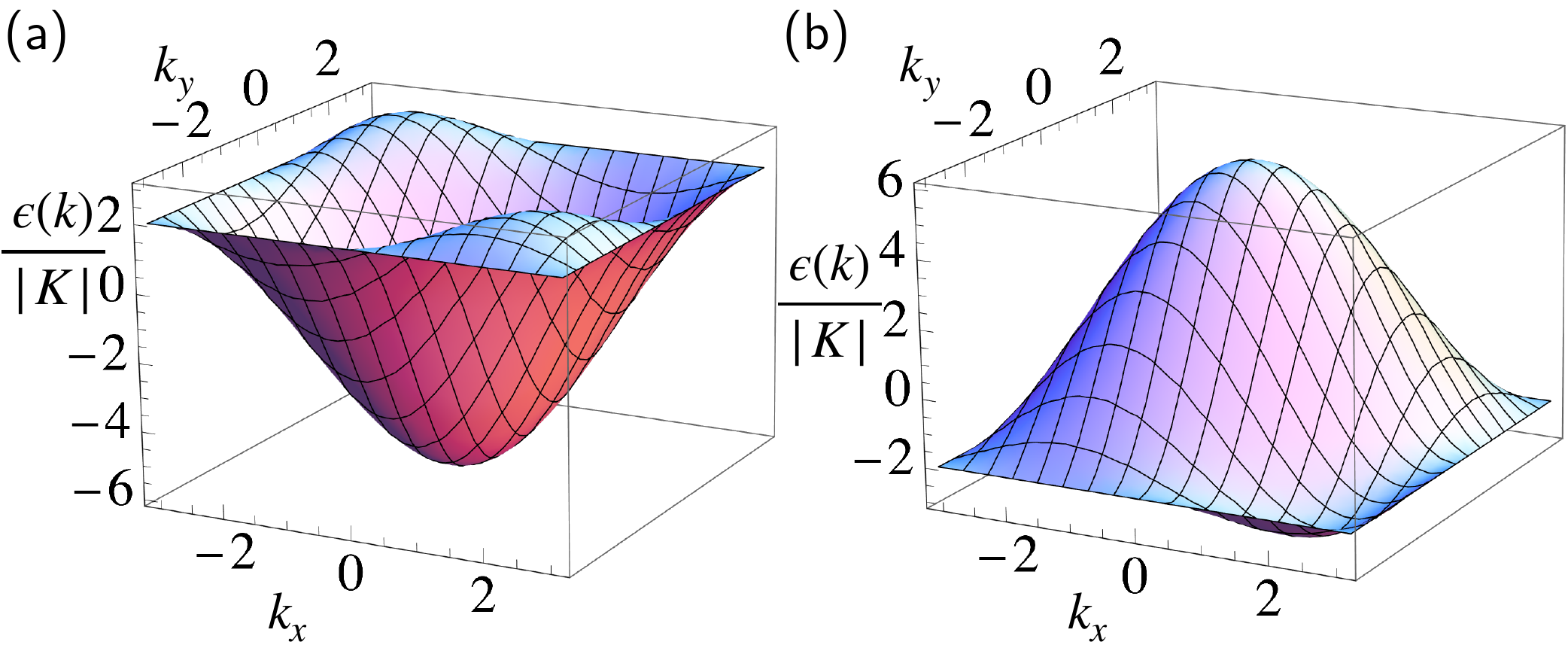}
\caption{
Number-conserving part of $\varepsilon$-particle bands
(on a square lattice) from $\tilde{H}$ with FM (a) and AFM
(b) couplings. The lattice constant is set to be $1$.
}\label{fig:epsilon-bands}
\end{figure}

By following the ideas of Ref.~\cite{Chen2022}, we will now show that the
vison in the FM Kitaev model is expected indeed to have the same phases as in the
trivial atomic insulator of spinons TC-AI$_0$, namely $\phiuc \teq 0$, whereas the vison
of the AFM Kitaev model has the same phase as in the trivial atomic insulator
TC-AI$_1$, i.e., $\phiuc \teq \pi$.
The $\varepsilon$-particle's band dispersion (obtained from the number-conserving
part of the $\varepsilon$-particle Hamiltonian) for $H$ with FM and AFM couplings
are shown in \Frefs{fig:epsilon-bands}(a) and (b) respectively.
For FM coupling, $\epsilon(0,0) \teq -6 |K|$,
$\epsilon(\pi, 0) \teq \epsilon(0, \pi) \teq \epsilon(\pi, \pi) \teq 2|K|$,
thus $\zeta_{(0,0)} \teq -1$ and $\zeta_k \teq +1$ for the rest HSPs.
Its parity indices at four HSPs are closer to those of TC-AI$_0$ state,
so we expect $\phiuc \teq 0$~\footnote{We have conjectured in Ref.~\cite{Chen2022}
that  states with only one non-trivial parity index have $\phiuc = 0$,
whereas states with $3$ non-trivial parity indices have $\phiuc = \pi$.
Our current numerics add further evidence to that conjecture.}.
As for AFM coupling, the $\varepsilon$ band dispersion is simply opposite to
that of the FM case. The parity indices $\zeta_{(0,0)} \teq 1$ and
$\zeta_k \teq -1$ at other HSPs, which is closer to that of TC-AI$_1$ state.
Therefore we expect $\phiuc \teq \pi$ from the arguments of Ref.~\cite{Chen2022}.

\begin{figure}
\centering
\includegraphics[width=0.48 \textwidth]{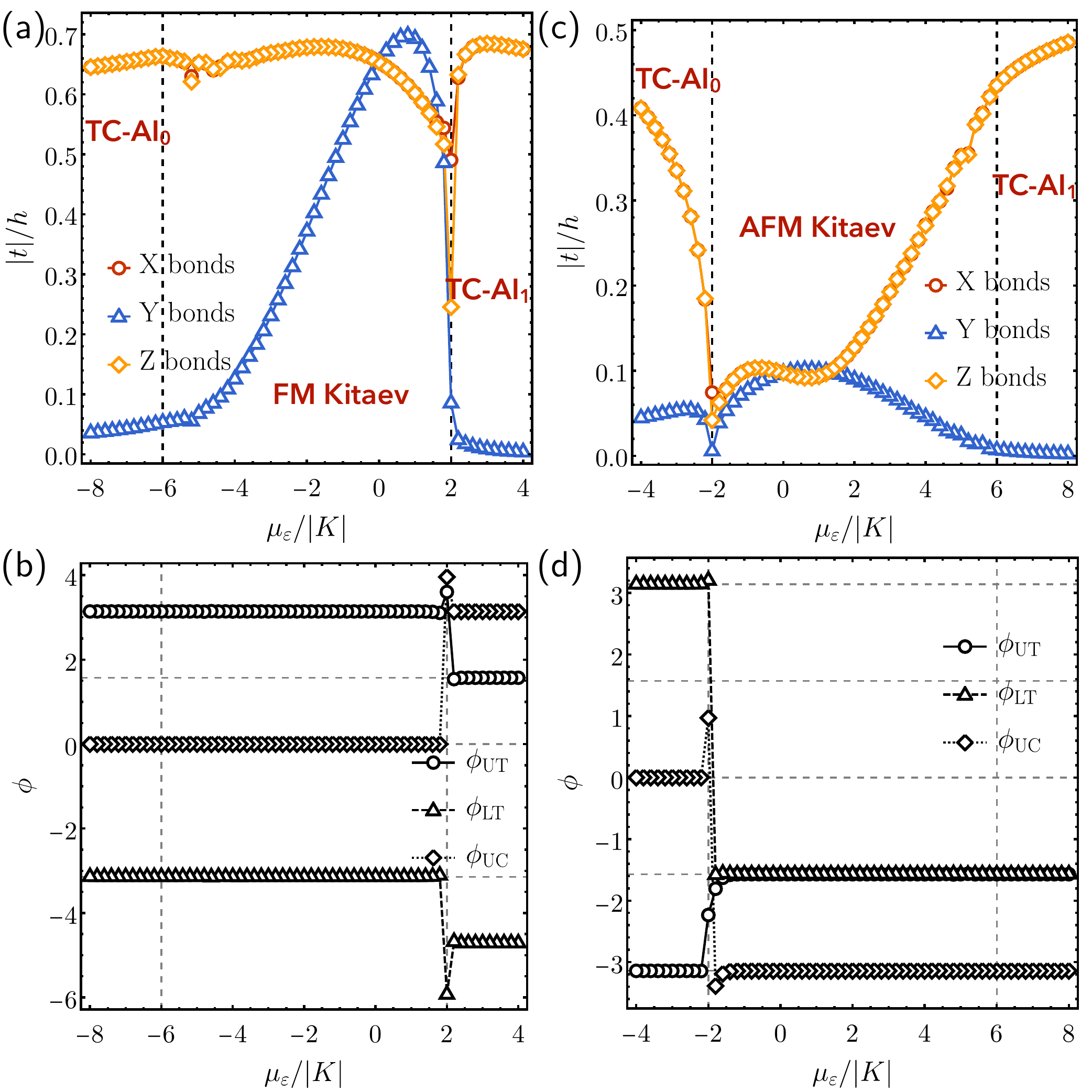}
\caption{
Vison hopping as a function of $\mu_\varepsilon$ at $\kappa/|K|$ \teq 0.1.
(a-b) The hopping amplitude and $\phi$ for FM coupling.
(c-d) Same quantities for AFM coupling.
Calculations are done with $N_\text{row} = 50$ and APBC.
}\label{fig:phi-mu}
\end{figure}

We will now provide direct numerical evidence for the above expected values
of $\phiuc$ in FM and AFM Kitaev model. To do so we add the following additional
term $H_1$ into $H$, which reads (after the unitary transformation $U$
introduced in \Secref{sec:duality}):
\begin{equation}\label{eq:H_1}
U H_1 U^{-1} = \sum_p \frac{\mu_\varepsilon}{2}
( \prod_{l \in \text{boundary}(p)} Z_l - 1 ).
\end{equation}
Here $p$ stands for a plaquette in the square lattice (see \Fref{fig:schematic}(b)).
Under the duality mapping discussed in \Secref{sec:duality} and in Ref.~\cite{Chen2022},
the above term is mapped into $\sum_p -\mu_\varepsilon c_p^\dagger c_p$.
Therefore, this term simply introduces a chemical potential to $\varepsilon$-particles
and therefore, it naturally drives the ground state into the commuting projector
limit of the TC-AI$_0$ state when $\mu_\varepsilon$ is sufficiently negative and into the one
of the TC-AI$_1$ state when $\mu_\varepsilon$ is sufficiently positive.
We calculated the phase that the vison acquires around a unit cell,
$\phiuc$ at different $\kappa$ and $\mu_\varepsilon$.
\Fref{fig:phi-mu} shows the result for $\kappa/|K| \teq 0.1$ with both
FM and AFM Kitaev couplings.
For FM coupling (\Frefs{fig:phi-mu}(a-b)), as $\mu_\varepsilon$ is tuned to the
band edges, the vison hopping crossing $Y$-bonds decreases while the hoppings crossing
$X$- and $Z$-bonds are approaching $\sim 0.7$.
When $-6|K|\le \mu_\varepsilon \tle 2 |K|$, $\zeta_{(0,0)} \teq -1$ and
$\zeta_k \teq 1$ for the other three HSPs, the parity indices' configuration
is closer to that of TC-AI$_0$ state. Indeed, the vison phase $\phiuc$ is always $0$
in this regime.
When $\mu_\varepsilon < -6 |K|$, since the bare $\varepsilon$ band is empty
($\zeta_k \teq 1$ at all four HSPs), the $\varepsilon$ ground state is
adiabatically connected to the TC-AI$_0$ state and $\phiuc \teq 0$.
Therefore, $\phiuc$ does note change its value across this FM-Kitaev to TC-AI$_0$ transition.
On the other hand, when $\mu_\varepsilon$ becomes larger than $2 |K|$, $\phiuc$ suddenly
jumps to $\pi$. This can be understood from the fact that in this regime, 
$\zeta_k \teq -1$ for all HSPs and the $\varepsilon$ ground state
is adiabatically connected to TC-AI$_1$, so $\phiuc$ should take the
same value as that of TC-AI$_1$ state.
For AFM coupling (\Fref{fig:phi-mu}(c-d)), as $\mu_\varepsilon$ is tuned to
the band edge, the vison hopping crossing $X$- and $Z$-bonds are close to
$\sim 0.5|h|$ while the hopping across $Y$-bonds decreases.
For $-2|K| \le \mu_\varepsilon \le 6 |K|$, $\zeta_{(0,0)} \teq 1$ while
$\zeta_k$ at all the other three HSPs are $-1$, which is closer to that of
TC-AI$_1$ state.
Indeed, we found that $\phiuc$ is always $\pi$ across the AFM-Kitaev to
TC-AI$_1$ ($\mu_\varepsilon \ge 6|K|$) transition.
On the other hand, when the $\varepsilon$ ground state enters the TC-AI$_0$ regime
($\mu_\varepsilon \tlt -2 |K|$), $\phiuc$ jumps to $0$.
Therefore, our results thus indicate that the vison phase $\phiuc$ for FM (AFM)
Kitaev coupling is indeed the same as that of the TC-AI$_0$ (TC-AI$_1$) state,
as expected from the considerations of Ref.~\cite{Chen2022}.

% ------------------------------------------------------------------------------
% FIGURE
% ------------------------------------------------------------------------------
% ------------------------------------------------------------------------------

% ------------------------------------------------------------------------------
% FIGURE
% ------------------------------------------------------------------------------
% ------------------------------------------------------------------------------
% 
% ------------------------------------------------------------------------------
% FIGURE
% ------------------------------------------------------------------------------
\begin{figure}
\centering
\includegraphics[width=0.45 \textwidth]{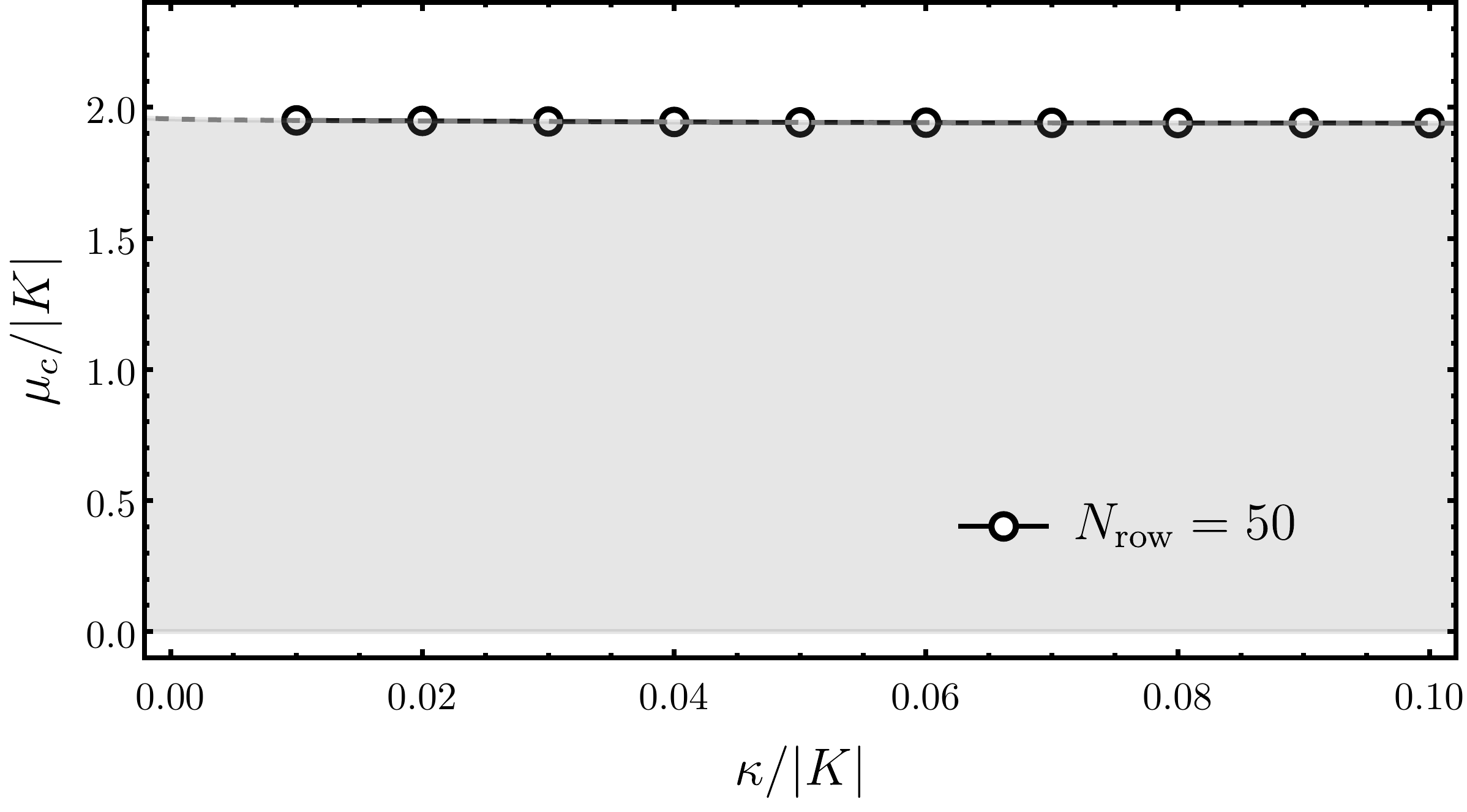}
\caption{
Critical $\mu_c$ for the $0$ to $\pi$ jump of $\phiuc$ at different $\kappa$
with FM coupling. $N_\text{row} \teq 50$. $\phiuc \teq 0$ in the gray region
with a caveat that $\phiuc$ at stricktly $\kappa \teq 0$ could be ill-defined.
}\label{fig:mu_c}
\end{figure}
% ------------------------------------------------------------------------------
% 
Moreover, we have performed same type of calculations at smaller $\kappa$ values, 
for each $\kappa$, we found a similar $0$ to $\pi$ jump behavior
at some ``critical'' chemical potential $\mu_c$.
\Fref{fig:mu_c} shows the result for $\mu_c$ at different $\kappa$ values
with a FM coupling. It can be seen that $\mu_c$ is not sensitive to $\kappa$ 
at small $\kappa$ values, this can be understood from the perspective of our
conjecture as $\kappa$ does not affect the
number-conserving part (hopping and chemical potential) of $\varepsilon$
particles' Hamiltonian, therefore $\zeta_k$ at all four HSPs are independent
of $\kappa$.
Below the $\mu_c$-$\kappa$ curve (the gray region in \Fref{fig:mu_c}),
there is always $\phiuc \teq 0$.
Note that although the extrapolation of finite $\kappa$ results
tends to suggest $\phiuc \teq 0$ for the $\kappa \teq 0$ FM Kitaev model
($\mu_\varepsilon \teq 0$), due to the reasons described
previously, $\phiuc$ at $\kappa \teq 0$ may not be a well-defined quantity.

\section{Summary and Discussion} \label{sec:discuss}
Using a recently developed exact duality mapping~\cite{Chen2022} that allows
to re-write the microscopic spin operators in terms of non-local visons and
fermionic spinons degrees of freedom,
we have investigated the nature of the motion of the emergent flux carrying
vison particles in the Kitaev honeycomb model perturbed by a Zeeman field.
This Zeeman field not only induces the well-known Haldane-type mass gap on
the itinerant Majorana fermions, but also induces vison hopping, breaking
the exact solvability of the model. 

We have seen that while the FM ($K<0$) and the AFM ($K>0$) Kitaev models
have the same Non-Abelian Ising topological order, they are sharply distinct phases
of matter when viewed as topologically ordered states that are enriched by the discrete
translational symmetry of the honeycomb lattice. As a consequence of this, the nature
of the motion of the vison particle is sharply distinct in the FM vs the AFM Kitaev
honeycomb models. For example, in the FM Kitaev model the vison acquires a trivial
phase when it encircles around a unit cell of the honeycomb Bravais lattice:
$\phiuc \teq 0$.
Since there is a single vison site per unit cell (which can be viewed as located
at the center of the plaquette), this implies that the vison moves effectively in
a single-site triangular lattice with zero flux. There is, therefore, a single vison
band which has zero Berry curvature and thus no associated intrinsic 
contribution the thermal Hall effect. To further back-up this conclusion, we have
shown that the vison phase around the unit cell remains unchanged across a phase
transition into another ground state that is adiabatically connected to the ground
state of the standard toric code model, which is a commuting projector Hamiltonian where
this phase can be computed fully analytically and where it is clear that translations
act non-projectively on its anyon excitations.
% A recent study reported a different
% value of the phase of the vison in the FM Kitaev model of $\phiuc \teq \pi$~\cite{Joy2021},
% but we believe this to be an artefact~\footnote{Possibly originating from the fact that
% this study did not include the Haldane mass on the majorana Fermions. We have shown
% that a small but non-zero Haldane mass is crucial to regularize the calculation of
% the vison phase, because in its absence there appear strong system size and boundary
% condition fluctuations.}.

On the other hand, in the AFM Kitaev model the vison acquires a non-trivial
phase when it hops around a unit cell: $\phiuc \teq \pi$.
As a consequence lattice translations are implemented projectively on the vison,
and the vison unit cell needs to be doubled. In this case, the vison has two
separate bands with non-zero Chern numbers $C=\pm1$, and an associated
contribution to the intrinsic thermal Hall effect.

There are also crucial energetic differences to the vison bands in the FM vs AFM models.
In the FM model, the magnitude of the vison hopping and band-width grow linearly
with the Zeeman field, $|t_\alpha|\approx 0.38 |h_\alpha|$, when the vison hops
across an $\alpha$-bond ($\alpha = X,Y,Z$) (see Fig.~\ref{fig:schematic})
at the leading perturbative order, which agrees with the value reported in
Ref.~\cite{Joy2021}.
For the AFM model our results are consistent with a vanishing
leading perturbative hopping of the vison that is linear in Zeeman fields, as also reported in Ref.~\cite{Joy2021}. However, we have seen that in the
presence of the Haldane mass term of the Majorana fermions,
$\kappa \approx h_x h_yh_z/K^2$~\cite{Kitaev2006}, the magnitude of the vison
hopping becomes non-zero and scales as:
$|t_\alpha| \propto |h_\alpha| |\kappa/K|^{\nu}$
when the vison hops across an $\alpha$-bond $\alpha \in \{X,Y,Z\}$
(see Fig.~\ref{fig:schematic}), with $\nu \approx 0.5$ at small $\kappa$
and $h_\alpha$. As a consequence of this, the visons in the FM model are substantiallly more
mobile at small values of the Zeeman field than in the AFM model.
Therefore the Zeeman field is expected to destabilize more easily the FM
Kitaev spin liquid via vison gap closing and condensation, relative to the
AFM Kitaev spin liquid. This is naturally consistent with a variety of numerical
studies which have reported that the FM Kitaev spin liquid is more fragile than the
AFM Kitaev spin liquid against Zeeman field
(see, e.g., Refs.~\cite{hickey2019emergence,zhu2018robust,gordon2019theory,gohlke2018dynamical}).
While the single-vison gap closing picture found here
provides a natural mechanism for an instability of the Kitaev spin liquid,
it should be noted that such a proliferation mechanism can also be applied to tightly
bounded vison pairs.
As discussed in Refs.~\cite{Zhang2021,Zhang2022}, a bounded vison pair also
gains dynamics under perturbations like Zeeman field, Heisenberg and Gamma interactions.
According to the magnitude of induced vison-pair hopping amplitudes of Refs.~\cite{Zhang2021,Zhang2022},
it was reported that the FM (AFM) Kitaev spin liquid is more robust (fragile) against Heisenberg
interaction and fragile (robust) against Zeeman field and Gamma interaction,
in agreement with our findings.

While the precise relation between the ideal spin liquids realized in the
weakly perturbed Kitaev model that we have studied and the possible spin
liquids observed in $\alpha$-RuCl$_3$ is currently far from clear, our study
highlights the crucial importance of the sign of the Kitaev coupling $K$ in
determining the universal properties of these states. While the larger share
of studies devoted to determining this sign have advocated it to be
ferromagnetic (see, e.g., Refs.~\cite{laurell2020dynamical,maksimov2020rethinking}
for summaries), direct experimental inference of this sign has heavily
relied in understanding the zig-zag AFM. However, in spite of being an ordered
state, the zig-zag AFM state is in itself still a highly quantum fluctuating
state that delicately depends on perturbations beyond the ideal Kitaev model~\cite{Winter_2017,rau2014trigonal,hickey2019emergence,zhu2018robust,gordon2019theory,gohlke2018dynamical,laurell2020dynamical,maksimov2020rethinking,kim2015kitaev,sorensen2021heart,gordon2019theory}.
One alternative state that is comparatively simpler to understand theoretically,
but which however remains less experimentally explored, is the high-field polarized state.
This state could offer a fresh alternative window to perform experiments that
could more confidently cement our knowledge of the sign of this important
coupling in $\alpha$-RuCl$_3$.

Finally, we would like to note that during the completion of our work an updated version of Ref.~\cite{Joy2021} appeared, and our results are in agreement with the various aspects where we overlap with that reference. The updated analysis of Ref.~\cite{Joy2021} was performed independently and largely in parallel to our work, and corrected an earlier version of that reference.

% \emph{Note}\;---\;%
%Finally, we would like to point out that our results for the phases of the
%vison in the FM model are different from
%those of Ref.~\cite{Joy2021}, which instead reported that $\phiut = \philt = \pi/2$ and
%$\phiuc = \pi$. This allowed the vison to have a projective implementation
%of translations with $\pi$-flux per unit cell, and consequently two Bands with non-zero Chern
%number and a finite contribution to the thermal Hall effect in the FM Kitaev model.
%We believe that this is an artefact  possibly arising because those calculations were
%performed with zero Haldane mass term for the itinerant Majorana fermions,
%which we have shown to be crucial in obtaining correct values of the vison hopping phases.
%This reference did not compute the phases in the AFM case since the vison hopping
%vanishes in this case with zero Haldane mass term.
%The authors of Ref.~\cite{Joy2021} have shared updated and unpublished calculations
%with us that include the Haldane mass term and where the vison phases acquired
%in the unit cells for both FM and AFM Kitaev models
%are now in agreement with our results.

% 
\acknowledgments
We would like to thank Peng Rao, Bernd Rosenow, Roderich Moessner, Alexander Tsirlin, Xue-Yang Song and T. Senthil for valuable discussions, and we specially thank Aprem Joy and Achim Rosch for several crucial discussions of their work that motivated our study. C.C. is supported by the Shuimu Tsinghua Scholar Program.

\appendix

\section{Vison hoppings with periodic boundary conditions} \label{sec:PBC}
\begin{figure*}
\centering
\includegraphics[width=0.8 \textwidth]{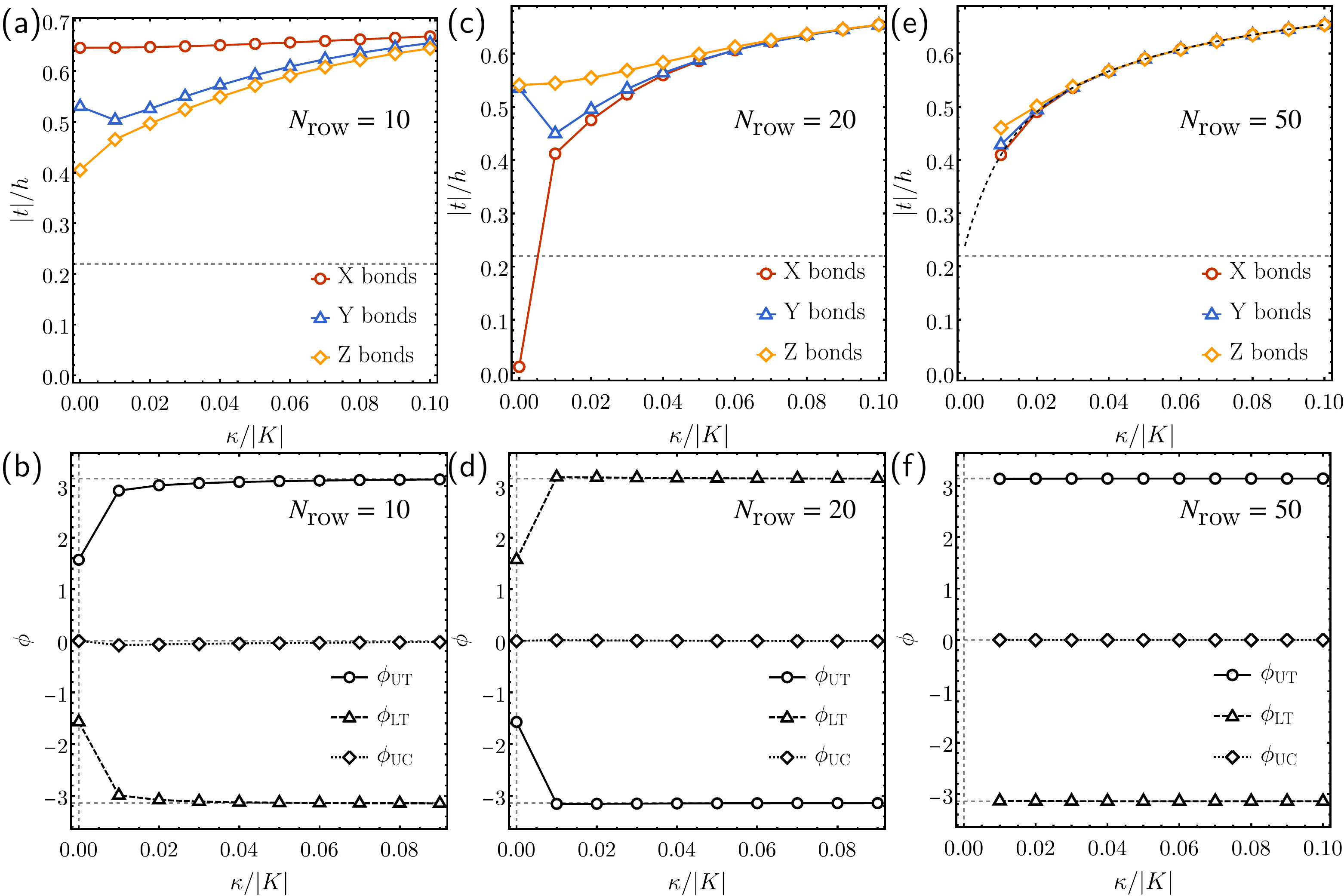}
\caption{
The vison hopping amplitude and $\phi$ with FM coupling at selected system sizes
with PBC for the spinons.
The results at finite $\kappa$ are consistent with APBC data presented in the
main text, whereas at $\kappa \teq 0$, the hoppings across certain bonds
can be dropped to very small values at,
which causes an ambiguity in determining the vison Berry phases.
}\label{fig:FM_vary_kappa-PBC}
\end{figure*}

\begin{figure*}
\centering
\includegraphics[width=0.6 \textwidth]{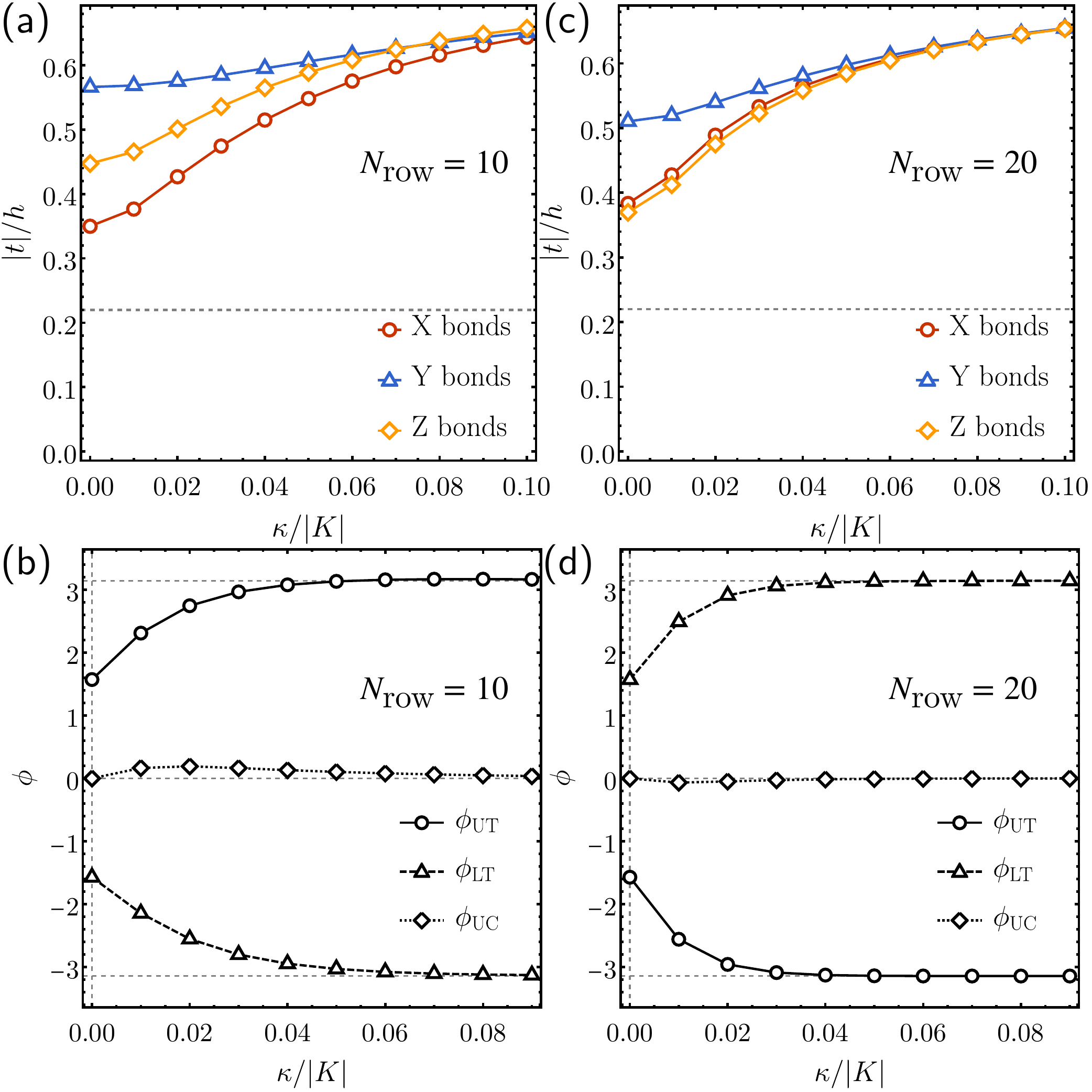}
\caption{
The vison hopping amplitude and $\phi$ with FM coupling at selected system sizes
with APBC for the spinons.
}\label{fig:FM_vary_kappa-APBC}
\end{figure*}

\begin{figure*}
\centering
\includegraphics[width=0.8 \textwidth]{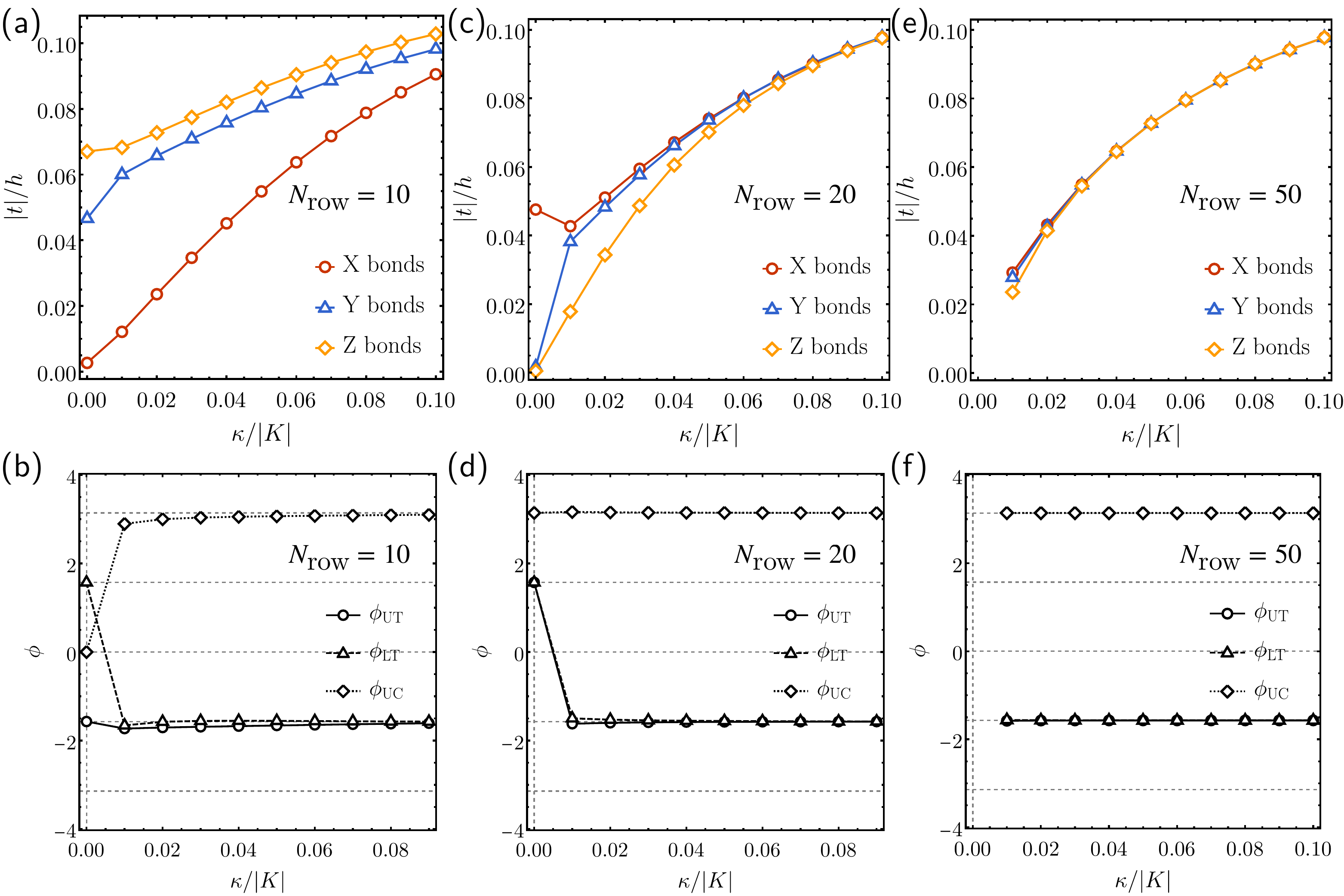}
\caption{
The vison hopping amplitude and $\phi$ with AFM coupling at selected system
sizes with PBC for spinons.
}\label{fig:AFM_vary_kappa-PBC}
\end{figure*}

\begin{figure*}
\centering
\includegraphics[width=0.6 \textwidth]{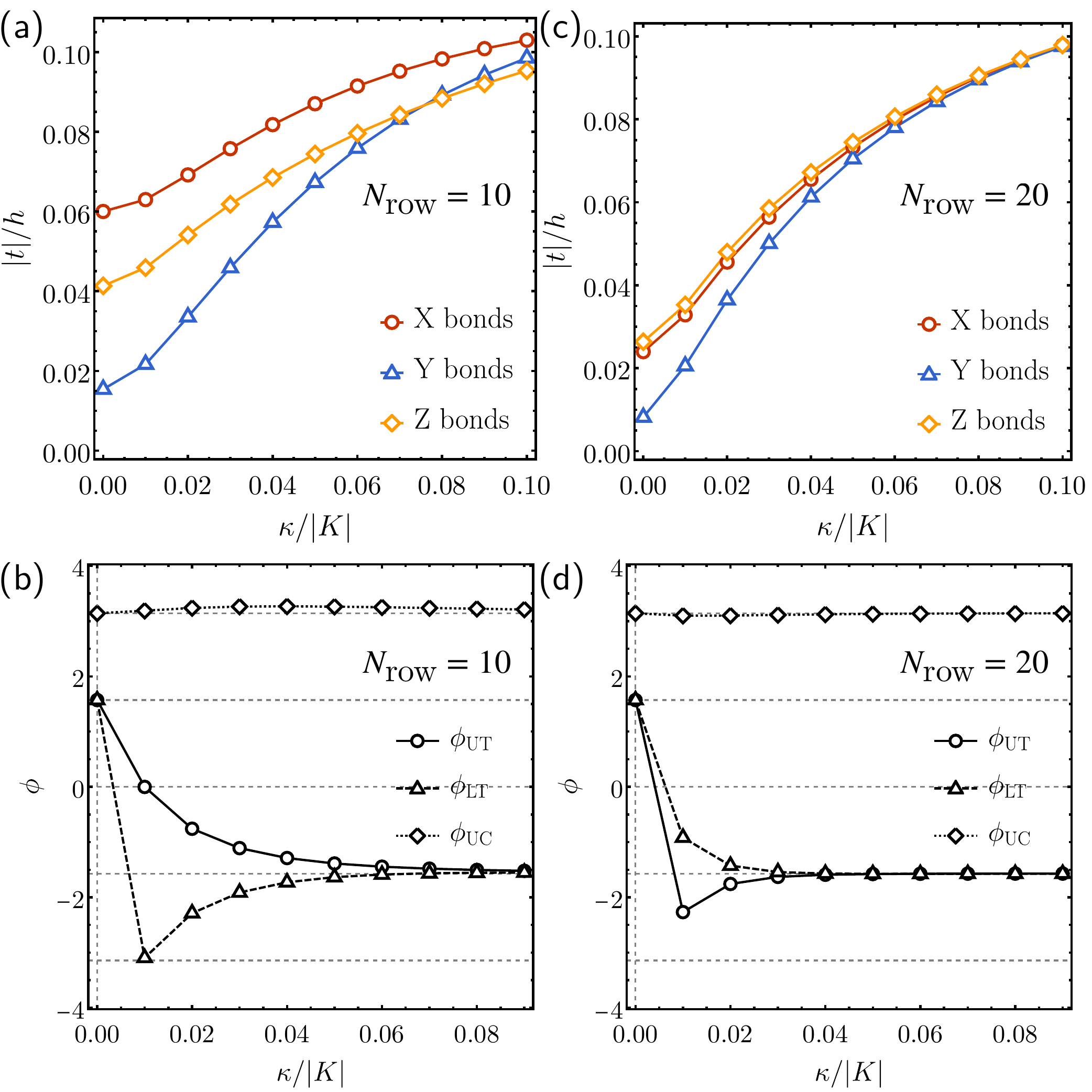}
\caption{
The vison hopping amplitude and $\phi$ with AFM coupling at selected system
sizes with APBC for spinons.
}\label{fig:AFM_vary_kappa-APBC}
\end{figure*}

In this section, we present the results of vison hoppings with
$(z_1, z_2) \teq (1, 1)$, i.e., PBC for $\varepsilon$ Majorana fermions.
The vison hopping and Berry phases with FM and AFM couplings are shown in 
\cref{fig:FM_vary_kappa-PBC,fig:AFM_vary_kappa-PBC} respectively.
At finite $\kappa$, the results are essentially the same as APBC results
discussed in the main text.
For comparison, we also present the APBC results at system size
$N_\text{row} \teq 10,\ 20$ with FM (\Fref{fig:FM_vary_kappa-APBC})
and AFM Kitaev coupling (\Fref{fig:AFM_vary_kappa-APBC}).

The Zeeman field is aligned along the $[111]$ direction
(same as in \Secref{sec:FM-AFM-hopping}) and we present the results
at $\kappa \ge 0$,
the Berry phases are obtained with positive $h$.
At finite $\kappa$, the vison hoppings and Berry phases are
independent of the boundary conditions
for both FM and AFM couplings (in the thermodynamic limit).
The main difference is at $\kappa \teq 0$.
In the case with APBC, the results for the system sizes presented
here seem to indicate $\phiuc = 0$ ($\pi$) for FM (AFM) couplings.
On the other hand, with PBC, the vison hoppings crossing
certain bonds could be very small,
e.g., see the data in \Fref{fig:FM_vary_kappa-PBC}(a),
\Fref{fig:AFM_vary_kappa-PBC}(a) and (c).
The smallness of the hopping across certain bonds creates
an obstacle to unambiguously identify the vison hopping phases.
For example, as shown in \Fref{fig:AFM_vary_kappa-PBC}(a), because the vison
hopping crossing certain bonds is close to zero, $\phiuc$ jumps from
$\pi$ to $0$ as $\kappa \rightarrow 0$.
In \Frefs{fig:AFM_vary_kappa-PBC}(c-d), although
$\phiuc \teq \pi$ at $\kappa \teq 0$, which looks consistent with the
APBC result, we caution that such a $\phiuc$ value is actually 
extracted from a numerically very small complex number
(both the real and imaginary part being $\sim 10^{-20}$).

In summary, according to our calculation, while the vison hopping at a finite
$\kappa$ is robust against spinon boundary conditions (Wilson loop sector),
the vison hopping at $\kappa \teq 0$ is very sensitive to the 
boundary conditions and system sizes.
We believe this is due to the gapless nature for the $\kappa \teq 0$
($B$-phase) Kitaev model. Our results then clearly indicate that a small but non-zero $\kappa$ is needed to properly regularize the vison hopping phases in the fully gapped topologically ordered state.

% ------------------------------------------------------------------------------
% FIGURE
% ------------------------------------------------------------------------------
% ------------------------------------------------------------------------------

% ------------------------------------------------------------------------------
% FIGURE
% ------------------------------------------------------------------------------
% ------------------------------------------------------------------------------

% ------------------------------------------------------------------------------
% FIGURE
% ------------------------------------------------------------------------------
% ------------------------------------------------------------------------------

% ------------------------------------------------------------------------------
% FIGURE
% ------------------------------------------------------------------------------
% ------------------------------------------------------------------------------

\section{Finite-size scaling of vison hoppings and associated phases} 
\label{sec:finite-size}
\begin{figure*}
\centering
\includegraphics[width=0.8 \textwidth]{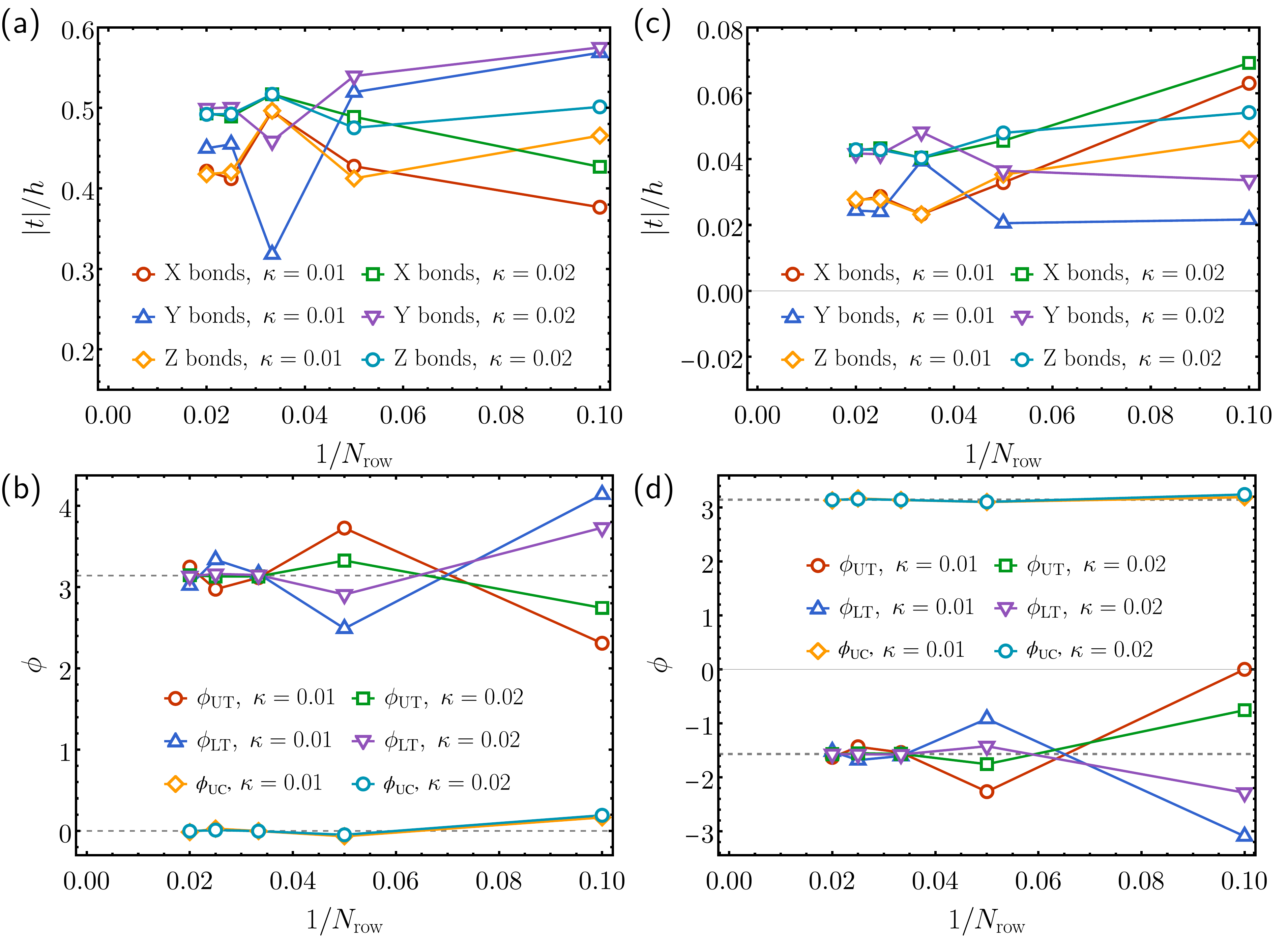}
\caption{
The scaling of vison hopping amplitudes and $\phi$ versus
$1/N_{\text{row}}$. (a-b) results with $K \teq -1$;
(c-d) results with $K \teq 1$.
}\label{fig:N-scaling}
\end{figure*}

Within our numerical calculation, the vison hoppings at finite $\kappa$
converges fast as the system size ($N_\text{row} \times N_\text{row}$)
increases.
Results at $\kappa \teq 0.01|K|$ and $\kappa \teq 0.02 |K|$ with APBC are presented
in \Fref{fig:N-scaling}.
Results with FM coupling are shown in \Fref{fig:N-scaling}(a-b)
and those with AFM coupling are shown in \Fref{fig:N-scaling}(c-d).

It was found that the convergence of vison hoppings
with repect to $N_\text{row}$ is faster at larger $\kappa$ values.
This also gives extra evidence of the strong sensitivity of the vison hoppings with system size in the case of strictly zero Haldane mass term $\kappa \teq 0$.

% ------------------------------------------------------------------------------
% FIGURE
% ------------------------------------------------------------------------------
% ------------------------------------------------------------------------------

\bibliography{reference.bib}

\end{document}